\documentclass[times,twocolumn,final,longtitle]{elsarticle}
\usepackage{jasr}
\usepackage{framed,multirow}
\usepackage{siunitx}
\usepackage{ulem}
\usepackage{amssymb}
\usepackage{latexsym}
\usepackage{amsmath}
\usepackage[switch]{lineno}
\usepackage{booktabs}
\usepackage{tabularx}
\usepackage{url}
\usepackage{xcolor}
\definecolor{newcolor}{rgb}{.8,.349,.1}

\usepackage[citebordercolor=white]{hyperref}
\usepackage{cleveref}
\usepackage{color} 
\usepackage{appendix}
\usepackage{threeparttable}
\journal{Advances in Space Research}

\begin{document}

\verso{D. Song \textit{etal}}

\begin{frontmatter}
\title{An Image Simulator of Lunar Far-Side Impact Flashes Captured from the Earth-Moon L2 Point}

\author{Da Song\texorpdfstring{\textsuperscript{a,b,c}, Hong-bo Cai\textsuperscript{b,$\ast$}, Shen Wang\textsuperscript{b},
Jing Wang\textsuperscript{a,b}}}

\ead{chb@nao.cas.cn}

\affiliation[1]{organization={Guangxi Key Laboratory for Relativistic Astrophysics, School of Physical Science and Technology, Guangxi University},
                addressline={100 East University Road},
                city={Nanning},
                postcode={530004},
                country={People’s Republic of China}}

\affiliation[2]{organization={Key Laboratory of Space Astronomy and Technology, National Astronomical Observatories, Chinese Academy of Sciences},
                addressline={A 20 Datun Road},
                city={Beijing},
                postcode={100101},
                country={People’s Republic of China}}
\affiliation[3]{organization={School of Astronomy and Space Science, University of Chinese Academy of Sciences},
                city={Beijing},
                country={People’s Republic of China}}
                
\received{$----$}
\finalform{$----$}
\accepted{$----$}
\availableonline{$----$}

\begin{abstract}
Impact flashes on the moon are caused by high-speed collisions of celestial bodies with the lunar surface. The study of the impacts is critical for exploring the evolutionary history and formation
of the Moon, and for quantifying the risk posed by the impacts to future human activity.
Although the impacts have been monitored from the Earth by a few projects in past 20 years, 
the events occurring on the lunar far side have not been explored systematically so far. 
We here present an end-to-end image simulator dedicated to detecting and monitoring the impacts from space, 
which is useful for future mission design. 
The simulator is designed for modularity and developed in the Python environment, which is mainly composed of four components: the flash temporal radiation, the background emission, 
the telescope and the detector 
used to collect and measure the radiation. Briefly speaking, with a set of input parameters, the simulator calculates
the flash radiation in the context of the spherical droplet model and the background emission from the lunar surface. The resulting images are then 
generated by the simulator after considering a series observational effects, including the 
stray light, transmission of the instrument, point spread function and multiple kinds of noise caused by a CCD/CMOS detector. The simulator is validated by comparing the calculation with the observations
taken on the ground. The modular design enables the simulator to be improved and enhanced
by including more complex physical models in the future, and to be flexible for other future space missions.

\end{abstract}

\begin{keyword}
\KWD Moon\sep Meteoroids\sep Impact Flash\sep Simulation
\end{keyword}
\end{frontmatter}


\section{Introduction}
\setlength{\baselineskip}{14pt}
\label{sec1}
The properties and motion  of small celestial bodies such as asteroids and comets are critical for understanding the evolutionary history of the solar system, because some of them are believed to contain the initial condition at the beginning of formation of the solar system \citep{davis2002collisional}.
Unlike the meteoroids observed on Earth, which generate bright light trails due to friction with the Earth's atmosphere, the small celestial bodies can directly impact on the lunar surface 
at high velocity without air resistance. The impact is predicted to produce detectable optical flashes (e.g., \cite{gordon1921meteors}). 
In addition to the properties and motion of the impactors, the study of the optical 
flashes enables us to explore not only the evolutionary history and formation processes of
the Moon, but also the potential threats posed by meteoroids to either spacecraft or astronauts.     
Furthermore, lunar dust is generated by meteoroid impacts, which prompted NASA to initiate the Lunar Atmosphere and Dust Environment Explorer (LADEE) mission \citep{elphic2011nasa}. As part of LADEE, the Lunar Dust Experiment (LDEX) has provided new insights into the lunar dust environment and its data can also be used to estimate the flux of impactors on the lunar surface \citep{horanyi2015lunar,pokorny2019meteoroids}.

The optical flashes attributed to meteoroid impacts on the Moon nearside have been reported
frequently by the ground-based observations in the past century. So far,  hundreds of impact events have been identified. For instance,  \cite{stuart1956photo} photographed a flash of light on the lunar surface likely attributed to a large meteoroid impact.
As the first unequivocal detection of impact on the night side of the Moon,
\cite{ortiz2000optical} reported unambiguous detection of five impact flashes during the 1999 Leonid meteor shower. Three of them have been seen simultaneously by other observers \citep[e.g.,][]{dunham1999lunar}. Four additional flashes attributed to an impact have 
been reported for the Leonid meteor shower occuring in November 2001 \citep{ortiz2002observation}.
In addition to the Leonid meteor shower, subsequent observations have detected lunar impact flashes from the Geminid \citep{yanagisawa2008lunar}, Lyrid \citep{moser2011luminous}, Perseid \citep{yanagisawa2006first}, and Taurid meteor showers \citep{cooke2006probable}, as well as sporadic impact flashes unrelated to meteor showers \citep{ortiz2006detection,cooke2007rate}.

In order to study lunar impact flashes systematically, \cite{cao2020lunar} listed several ground-based monitoring programs focusing on the near side of the Moon that have been established since 2005, including the NASA Lunar Impact Monitoring Program \citep{suggs2008nasa} initiated in 2006, the Moon Impacts Detection and Analysis System (MIDAS) program \citep{madiedo2010robotic} started in 2009, and the NEO Lunar Impacts and Optical Transients (NELIOTA) program \citep{bonanos2015neliota} initiated in 2015. By using a 
1.2-meter telescope equipped with a scientific camera (Andor Zyla 5.5 sCMOS) \citep{bonanos2018neliota,xilouris2018neliota,liakos2019neliota}, NELIOTA had confirmed 187 flash events\footnote{https://neliota.astro.noa.gr/} until July 2023. 

It has still not been achieved to monitor the impact flashes occurring on the far side of the Moon.
To address this, various research institutions globally have initiated studies on lunar far side impact flashes.  One notable effort is the Lunar Meteoroid Impacts Observer (LUMIO) mission \citep{cervone2022lumio}. This CubeSat mission began its Phase 0 study in 2017, followed by an independent mission assessment by European Space Agency (ESA) in 2018. 
After successfully completing Phases A and B, ESA approved LUMIO to advance to Phases C and D in June 2024, which includes hardware development, with a potential launch as early as 2027.
Observations of these impacts would allow us to examine the properties of lunar soil on the far side and to identify the risk caused by the impacts to future human activity on the far side.

In this paper, we develop an end-to-end image simulator for a proposed mission 
that monitors the lunar far side impact flashes from the Earth-Moon L2 point. 
With the simulated images with a set of instrumental parameters, 
the simulator is useful for future mission design and optimization. 
The paper is organized as follows. Section 2 describes the conception of the 
proposed mission. 
Based on the conception, an end-to-end image simulation is formulated in Section 3.
The results of the simulation is shown in Section 4.

\section{Mission Concept}

The current image simulator is developed for a proposed mission that is described briefly as follows. By pointing to the shadow region of the Moon, a camera
working simultaneously in multiple optical bands is used to monitor the 
far side of the Moon from the Earth-Moon L2 point that is roughly 65,000 kilometers from the Moon \citep{burns2013lunar}. 
In each band, either a CCD or a CMOS is adopted as the detector. LUMIO will operate in a quasi-halo orbit around the Earth-Moon L2 with a CCD, collecting data in the visible (VIS) and near-infrared (NIR) bands.
An impact flash will be identified in real time from the obtained images 
by onboard dedicated pipelines based on traditional astronomical methods. 

Our simulator has some similarities with the work of \cite{merisio2023present,topputo2023meteoroids}, particularly in simulating lunar impact flashes and evaluating camera performance using metrics like signal-to-noise ratio (SNR).  However, we have further advanced these efforts.  First, our simulator generates images that show the visual process of the impact flash, while their model focuses on statistical analysis based on calculations.  Additionally, our model simulates how the flash dims over time until it disappears, providing a more detailed timeline for analysis.  In contrast, their study uses an average flash temperature for evaluation.  By expanding on this research, our simulator offers a new perspective on analyzing the time evolution of flash events.

\section{An End-to-end Image Simulator}

Based on the mission concept described above, an end-to-end image simulator
is developed and specified in this section.  

\begin{figure*}
\centering
\includegraphics[scale=0.6]{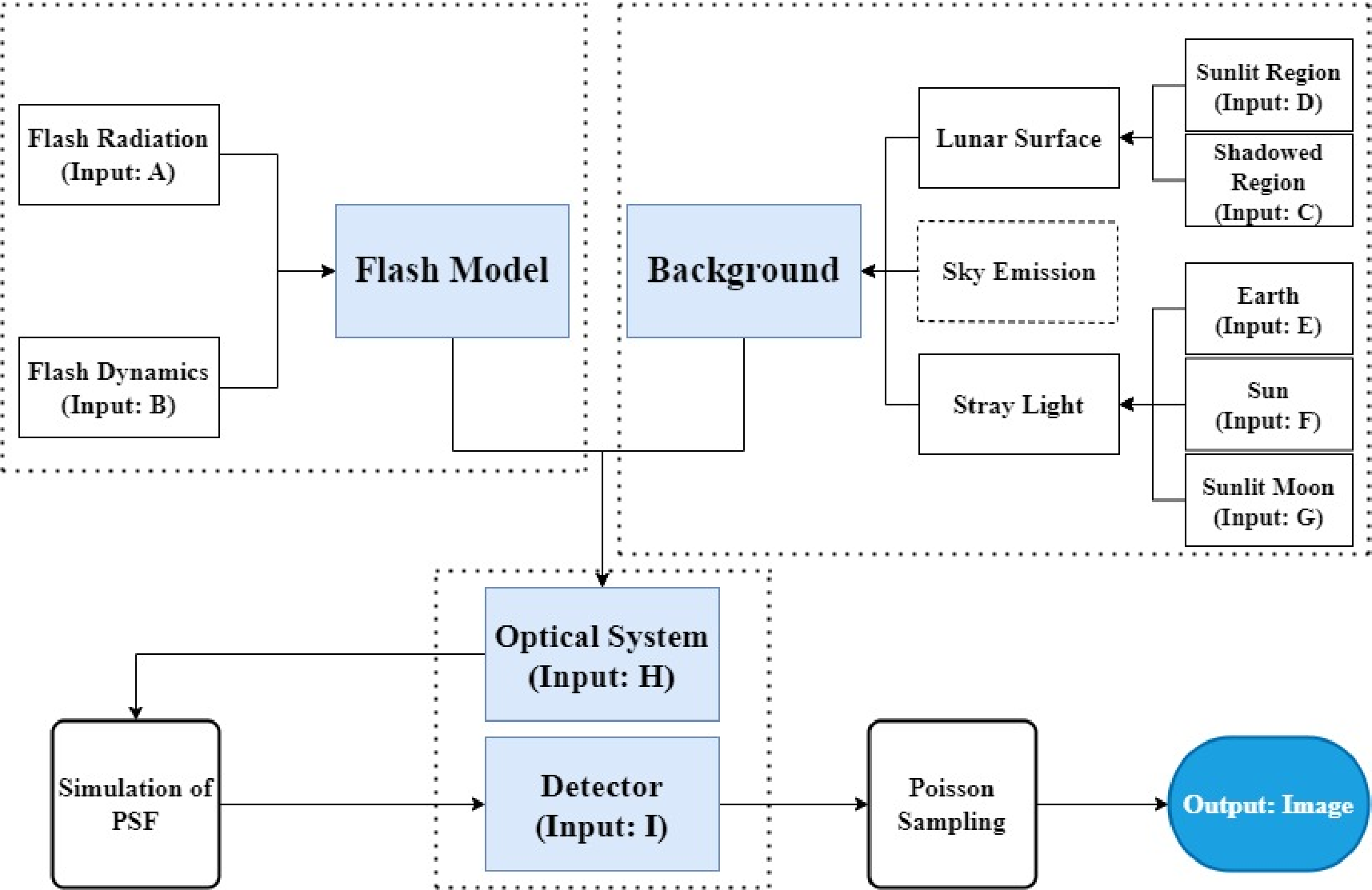}
\caption{Flow diagram of the flash image simulator developed by us. There are four modules in the simulator, i.e., the flash model, background, optical system and detector. In the background module, the sky emission is not included in the current version, although an interface is reserved.}
\label{figdiagram}
\end{figure*}

\autoref{figdiagram} illustrates the design of the current simulator. Briefly speaking, 
the simulator is mainly composed of four modules: flash, background, optical system and 
detector. With a set of input parameters listed in Table~\ref{tab:table1}, the simulator calculates the emission of a given
impact flash, along with the background contributed by various sources. The emission is then 
translated to Analog-to-Digital Unit (ADU) counts from the detectors in the focal plane by the optical system module and detector module, which  
results in a final image as an output of the simulator. The details of 
each of the modules are described as follows.

\begin{table*}[htbp]
\centering
\caption{The list of the input parameters required for the simulator.}
\label{tab:table1}
\begin{tabular}{ccc}
\toprule
Parameters  & Unit &  Description \\
\midrule
\multicolumn{3}{c}{Group A}\\
\midrule
$T_0$  & K & Peak temperature of the flash\\
\midrule
\multicolumn{3}{c}{Group B}\\
\midrule
$m$ & g & Mass of meteoroid\\
$m_\rho$ & $\mathrm{g\ cm^{-3}}$ &Density of meteoroid\\
$\upsilon$ & $\mathrm{km\ s^{-1}}$ & Impact velocity of meteoroid\\ 
$R_{\mathrm{d}}$ &  $\mu$m & Radius of individual molten droplet\\
\midrule
\multicolumn{3}{c}{Group C}\\
\midrule
$T_{\text{shadow}}$ & K & Lunar shadow temperature\\
$\bar{\varepsilon}$ & / & Average emissivity of the lunar shadow region\\
\midrule
\multicolumn{3}{c}{Group D}\\
\midrule
$f_{\odot}$ & $\mathrm{W\ m^{-2}}$ & Solar irradiance\\
$\eta$ & / & Average albedo of the Moon\\
$\mathbf{s}$ & / & Direction of the solar light\\
\midrule
\multicolumn{3}{c}{Group E}\\
\midrule
$f_{\odot}$ & $\mathrm{W\ m^{-2}}$ & Solar irradiance\\
$\eta_{\text{E}}$ & / & Average albedo of the Earth\\
\midrule
\multicolumn{3}{c}{Group F}\\
\midrule
$f_{\odot}$ & $\mathrm{W\ m^{-2}}$ & Solar irradiance\\
\midrule
\multicolumn{3}{c}{Group G}\\
\midrule
$f_{\odot}$ & $\mathrm{W\ m^{-2}}$ & Solar irradiance\\
$\eta$ & / & Average albedo of the Moon\\
\midrule
\multicolumn{3}{c}{Group H}\\
\midrule
$R$ & $\mathrm{m}$ & The distance between flash and observer\\
$D$ & $\mathrm{mm}$ & Aperture\\
$d_f$ & $\mathrm{mm}$ & Focal length\\
$\tau_{t}$ & / & Transmittance of the optical system\\
$\bar{\lambda}$ & $\mathrm{m}$ & effective wavelength\\
$\lambda_1$ & $\mathrm{m}$ & upper wavelength limit\\
$\lambda_2$ & $\mathrm{m}$ & lower wavelength limit\\
$PST$ & / & Stray light suppression coefficient of the optical system\\
\midrule
\multicolumn{3}{c}{Group I}\\
\midrule
$N$ & / & The number of active pixels\\
$d_{\text{pix}}$ & $\mu$m & Individual pixel size\\
$\Delta t$ & seconds & Exposure time\\
$\eta_{q}$ & / & Average quantum efficiency of the detector\\
$N_{\text{r}}$ & e$^{-}$ & Readout noise\\
$I_{\text{d}}$ & e$^{-}$pixel$^{-1}$s$^{-1}$ & Dark current\\
$G$ & / & Gain\\
\bottomrule
\end{tabular}
\end{table*}

\subsection{Emission from Lunar Impact Flash}
We simulate the emission and evolution of an impact flash by following the model given in Yanagisawa \& Kisaichi (2002). In the model, the emission of a flash can be well described by a blackbody at a temperature of $T$
\begin{equation}
\centering
f_{\text{flash}}(\lambda,T) = \frac{2\pi hc^2}{\lambda^5} \cdot \frac{1}{\mathrm{e}^{\frac{hc}{\lambda k_\mathrm{B}T}} - 1}
\label{eq1}
\end{equation}
where $h$,$c$ and $k_{\mathrm{B}}$ are the Planck
constant, the speed of light and the Boltzmann constant, respectively. 

With the blackbody emission, 
we calculate the flux of an impact flash within a wavelength range from 
$\lambda_1$ to $\lambda_2$ by an integration 
\begin{equation}
\centering
F(t) = \int_{\lambda_1}^{\lambda_2} \frac{f_{\text{flash}}(\lambda,T(t)) \cdot S_{\mathrm{e}}}{f \pi R^2} \, \mathrm{d}\lambda
\label{eq2}
\end{equation}
where $S_\mathrm{e}$ is the effective area of the flash, $f=4$ since we adopt the molten droplet model, which is assumed to be standard spherical radiation, and $R$ the distance to an observer.

\subsubsection{Flash Cooling}

By following the model in \cite{yanagisawa2002lightcurves}, the thermal evolution of the molten droplet can be described by the heat conduction equation:
\begin{equation}
\centering
\frac{\partial T}{\partial t} = \frac{\kappa}{r^2}\frac{\partial}{\partial r}\left( r^2 \frac{\partial T}{\partial r} \right)
\label{eq3}
\end{equation}
with a radiative boundary condition at the surface with a temperature of $T_{\mathrm{s}}$:
\begin{equation}
\centering
k\frac{\partial T}{\partial r} = -\sigma T_{\mathrm{s}}^4
\label{eq4}
\end{equation}
where $\sigma$ is the Stefan–Boltzmann constant, $\kappa$ is the thermal diffusivity given by $\kappa=k/(c_\mathrm{p}\rho)$, $k$ the thermal conductivity, $\rho$ the density of the droplet, and $c_\mathrm{p}$ the specific heat.

The cooling model is involved in our simulator in a simplified way by assuming a uniform temperature with the droplet. This approximation is reasonable because the thermal conduction is usually faster than the radiation process \citep[e.g.,][]{bouley2012power}. 
In this scenario, the cooling of the droplet can be approximated by the equation integrated over volume:
\begin{equation}
\centering
\frac{4}{3} \pi R_{\mathrm{d}}^3 \rho c_{\mathrm{p}} \frac{\mathrm{d}T}{\mathrm{d}t} = -4\pi R_{\mathrm{d}}^2 \sigma T^4
\label{eq5}
\end{equation}
where $R_{\mathrm{d}}$ is the droplet radius being typically on the order of tens to hundreds of micrometers \citep{mckay1991lunar}. This equation has an analytical solution:
\begin{equation}
\centering
T(t) = \frac{T_0}{\left(1 + \frac{9t\sigma T_0^3}{\rho R_{\mathrm{d}} c_\mathrm{p}}\right)^{\frac{1}{3}}}
\label{eq6}
\end{equation}
where $T_0$ is the peak temperature. 
The typical value of $c_{\mathrm{p}}= 1.3\ \mathrm{J\cdot g^{-1}K^{-1}}$ 
(i.e., solid glass) is adopted in the simulator for the droplet \citep{cintala1992impact}.

\subsubsection{Flash Effective Area}
According to the spherical droplet model proposed by {\cite{yanagisawa2002lightcurves}, the 
effective area of the flash ($S_{\mathrm{e}}$) can be estimated from the liquid volume $V$
\begin{equation}
\centering
S_{\mathrm{e}} = \frac{V}{\frac{4}{3} \pi R_{\mathrm{d}}^3} \cdot 4 \pi R_{\mathrm{d}} ^2 =\frac{3V}{R_{\mathrm{d}}}
\label{eq7}
\end{equation}
The volume $V$ is related with the mass ($m$), density ($m_\rho$), and impact velocity ($\upsilon$) of 
the meteoroid as 
\begin{equation}
\centering
V = \frac{m}{m_\rho} (c_1 + c_2 \upsilon + c_3 \upsilon^2)
\label{eq8}
\end{equation}
where $c_1$, $c_2$ and $c_3$ are constants depending on the meteoroid density and 
lunar regolith temperature, as detailed in \cite{cintala1992impact}. In the simulations below, these constants are as follows: for the molten droplet model, $c_1$ = -12.1, $c_2$ = 1.69, and $c_3$ = 0.0233; for the vapor model, $c_1$ = -0.657, $c_2$ = -0.107, and $c_3$ = 0.0211.

\subsection{Background Emission}

The background signal from the lunar surface is taken into account in the current simulator. 
The emission and reflection of the lunar surface is calculated at different phases after taking the direction of the sunlight into account. The simulated brightness of the lunar surface at different phases is illustrated in \autoref{fig2}.

\subsubsection{Surface emission from Shadow Region}

The emission from the shadow region is simulated by its own thermal radiation, which is controlled by the surface temperature and the emissivity of the regolith. The thermal radiation per unit area within the wavelength range $\lambda_1$ to $\lambda_2$, which is typically considered to be uniformly emitted over a hemispherical area from the lunar surface, is given by:
\begin{equation}
\centering
f_{\text{shadow}} = \int_{\lambda_1}^{\lambda_2} \frac{2\pi hc^2}{\lambda^5} \frac{1}{\mathrm{e}^{hc/(\lambda k_B T_{\text{shadow}})} - 1} \bar{\varepsilon} \, \mathrm{d}\lambda
\label{eq9}
\end{equation}
where $T_{\text{shadow}}$ is the surface temperature of the shadowed region of the Moon, approximately 110 K \citep{vaniman1991lunar}, and $\bar{\varepsilon}$ is the average emissivity of the lunar regolith over the wavelength range $\lambda_1$ to $\lambda_2$.
While this equation assumes uniform thermal radiation emission, the received radiation is related to the observation angle. Specifically, when observing at higher angles (relative to the surface normal), less radiation is received compared to lower angles, even if the surface radiates isotropically.

\subsubsection{Surface reflection from Sunlit Region}

We simulate the radiation signal from the sunlit region of the lunar surface by a reflection of the solar radiation in the traditional Lambert illumination model \citep{lambert1760photometria} : 
\begin{equation}
    f_{\mathrm{sunlit}} = \int_{\lambda_1}^{\lambda_2} f_{\odot} \cdot \eta\cdot\cos{\theta}\mathrm{d}\lambda
\end{equation}
where $\eta$ is the albedo of the lunar surface, which is taken as 0.15 and can be referenced in \cite{heiken1991lunar,muinonen2011lunar}, $f_\odot$ the solar irradiance at the surface,
and $\theta$ the angle between the reverse vector of the incident light and the normal vector of the surface.
Our simulation adopts the irradiance outside the Earth atmosphere given by the World Meteorological Organization Instrument and Observing Methods Committee at its eighth session\footnote[2]{WMO: \href{https://library.wmo.int/idurl/4/41712}{https://library.wmo.int/idurl/4/41712}}. For the $R-$band (550-800 \text{nm}), the solar irradiance is 377 $\mathrm{W/m^2}$, and for the $I-$band (700-950 \text{nm}), it is 271 $\mathrm{W/m^2}$.

\subsection{Image}
\subsubsection{Flash}

With the flux of a given impact $F$ estimated in Eq (2), the total number of electrons 
generated on the CCD within an exposure time of $\Delta t$ can be estimated as:
\begin{equation}
\centering
    N_{s}=\frac{\pi D^{2} \cdot F \cdot \tau_{t} \cdot \eta_{q}\cdot\Delta t}{4E_{0}} \label{eq11}
\end{equation}
where $E_0= hc/\bar{\lambda}$ is the energy of a single photon with an effective wavelength of $\bar{\lambda}$, $D$ the effective aperture of the telescope. $\tau_{t}$ and $\eta_{q}$ are the total throughput of the optical system and the average quantum efficiency of the detector, respectively.

In our simulator, each flash is considered as a point source whose profile at the focus 
is described by the point spread function (PSF). In the CCD coordinate $(x, y)$, the PSF 
is specified as a 2-dimensional Gauss function:  
\begin{equation}
\centering
G(x,y) = G_0 \mathrm{e}^{-\frac{(x-x_0)^2}{2\sigma_x^2} - \frac{(y-y_0)^2}{2\sigma_y^2}}
\label{eq23}
\end{equation}
where $G_0$ is the maximum value at the center, $\sigma_x$ and $\sigma_y$ are the dispersion 
in the $x$ and $y$ directions, respectively.

\subsubsection{Background}

According to the estimated background emission, the corresponding photon-generated electrons recorded by 
each CCD pixel is calculated as 
\begin{equation}
\begin{aligned}
N_b &= \frac{f_j \cdot \left(\frac{d_{\text{pix}}^2}{d_f^2}\cdot R^2\right)  \cdot \frac{\pi D^2}{4} \cdot \tau_t \cdot \eta_q \cdot \Delta t}{2 \pi R^2 \cdot E_0} \\
&= \frac{f_j \cdot d_{\text{pix}}^2 \cdot D^2 \cdot \tau_t \cdot \eta_q \cdot \Delta t}{8 d_f^2 \cdot E_0}
\end{aligned}
    \label{eq13}
\end{equation}
The term \(\frac{d_{\text{pix}}^2}{d_f^2} \cdot R^2\) represents the sky area corresponding to a single pixel at a distance \(R\). \(\frac{\pi D^2}{4}\) indicates the effective area of the telescope's aperture. \(2\pi R^2\) represents the total emission area of a hemisphere.
\begin{itemize}
  \item \( f_j \) represents the radiative flux within the shadowed or sunlit regions of the Moon.
  \item \( d_{\text{pix}} \) denotes the size of each pixel.
  \item \( D \) is the effective aperture of the telescope.
  \item \( \tau_t \) and \( \eta_q \) denote the total transmission of the optical system and the average quantum efficiency of the detector, respectively.
  \item \( \Delta t \) is the exposure time.
  \item \( d_f \) represents the focal length of the optical system.
  \item \( E_0 \) is the energy of a single photon.
\end{itemize}

\subsubsection{Stray Light}
The pollution produced by stray light can reduce the signal-to-noise ratio of the target by
enhancing the background level artificially. 

In ground-based observations, stray light is often significantly higher near the Moon’s terminator and less pronounced in regions farther away.

For space-based observations, the distribution of stray light may vary due to different environmental factors and the spacecraft's design. The spacecraft's attitude, position, and orientation can all influence stray light distribution. For example, the spacecraft’s solar panels or other components might reflect light into the telescope, creating stray light.   Additionally, the spacecraft might cause occlusions or shadows that affect the distribution of light pollution.  These factors and design considerations represent current limitations of the simulator. Consequently, at this stage of the simulation, stray light is assumed to be uniformly distributed in the focal plane.

The number of photon-generated electrons of each pixel can be estimated as 
\begin{equation}
\centering
N_n = \frac{\sum f_i \cdot \mathrm{PST}_i \cdot d_{\text{pix}}^2 \cdot \eta_q \cdot\Delta t}{E_0}
\label{eq14}
\end{equation}
where $f_i$ is the flux of sunlight, the illuminated part of the Earth, or the illuminated part of the Moon at the entrance pupil of the telescope. Specifically, for the Sun, we use the previously mentioned \( f_{\odot} \). For the Earth and Moon, the flux is expressed as 
\begin{equation}
f_i = \frac{f_{\odot} \cdot S_i \cdot \eta_i}{2 \pi R_i^2}
\end{equation}
where \( \eta_e = 0.29 \) is the average albedo of the Earth \citep{stephens2015albedo}, and \( R_i \) denotes the distances from the observation point to the Earth and the Moon, respectively. The illuminated area on one side of the Earth and Moon is given by \( S_i = 2 \pi r_i^2 \cdot A \), with \( r_i \) representing the radii of the Earth and the Moon. \( A \) represents the ratio of the illuminated area to the total area on that side (depending on the moon phase). $E_0$ the energy of a single photon, $\eta_q$ the quantum efficiency, $d_{\mathrm{pix}}$ the size of each pixel, and $\Delta t$ the exposure time. 
PST (a short for point source transmittance), defined as 
\begin{equation}
 \mathrm{PST}(\alpha) = \frac{E_\mathrm{d}(\alpha)}{E_{\mathrm{s}}(\alpha)}
\end{equation}
is used to characterize the stray light rejection capability of the optical system.
Here, $E_\mathrm{d}(\alpha)$ and $E_\mathrm{s}(\alpha)$ are the irradiance received 
at the focal plane and that from an off-axis point source.
It is influenced by the materials and structures of optical components such as baffle. A smaller PST value indicates lower levels of stray light received by the image sensor. PST is closely related to the incident angle $\alpha$ of the source and drops significantly with the angle.

\subsubsection{Detector}

A set of parameters is adopted in our simulator to model the performance of a perfect 
CCD detectors \citep{konnik2014high}. 
In addition to $\eta_q$ and $d_{\text{pix}}$ used above, the parameters include the total pixel 
numbers $N\times N$, bias level, dark current, readout noise and CCD gain $G$.  

The field-of-view is therefore inferred to be $\Delta\Omega=(N\alpha_p)^2$, where $\alpha_p$ is the 
angular size of a single pixel in the sky:
\begin{equation}
\centering
\alpha_p = \frac{d_{\text{pix}}}{d_f} \cdot \frac{180}{\pi} \ \mathrm{degree}
\end{equation}
Note that the approximation \(\frac{d_{\text{pix}}}{d_f}\) is valid only for small angular sizes.

Imperfections of the CCD, such as  photo-response non-uniformity, as well as the CCD smear effect, are ignored in our 
current simulator, although it can be easily implemented in the simulator in the future. 

\subsubsection{Noise and Final Image}
The noise generated during observations mainly consists of inherent noise and signal noise.  
The inherent noise originates from the detector, including readout noise and dark current noise.  The signal noise is mainly caused by the quantum nature of the signal and background emission.

With the calculated electrons from the flash, background and detector, the resulting image is finally obtained
by a statistical sampling in each pixel according to the traditional Poisson distribution
\begin{equation}
   P(k)=\frac{\mathrm{e}^{-N}N^k}{k!} 
\end{equation}

After the sampling, the final ADU of each pixel is obtained from the total electrons by a given $G$
\begin{equation}
    \mathrm{ADU} = \frac{N_\mathrm{e}}{G}
\end{equation}
where $N_{\mathrm{e}}$ is the number of electrons generated by the incoming photons.

\begin{figure}
\centering
\includegraphics[width=0.5\textwidth]{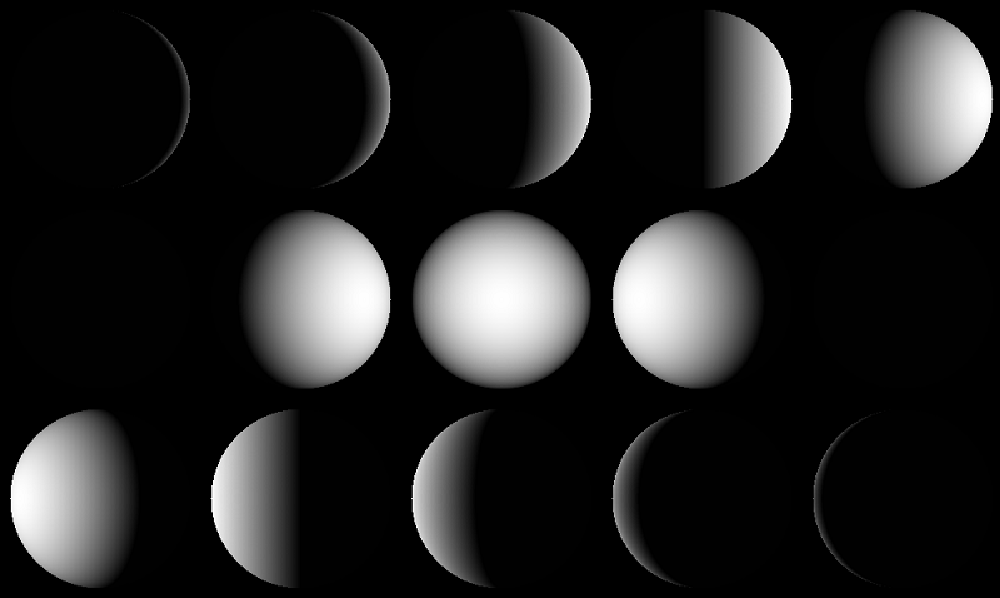}
\caption{The simulated lunar surface brightness at different phases. See Section 3.2 for the details.}
\label{fig2}
\end{figure}

\section{Results and Examples}

By following the methods described in Section 3, a simulator is developed with the 
Python 3.9 software in the Windows 10 operating system. 
The packages of NumPy \citep{harris2020array}, SciPy \citep{virtanen2020scipy}, and Astropy \citep{robitaille2013astropy} are required for running the simulator, which we will make available for download to readers on GitHub\footnote[3]{https://github.com/luckydog9?tab=repositories}}. 
To generate a series of images, the simulator inputs a set of parameters describing the flash radiation, the observation equipment, and the observation conditions, such as distance from an observer, moon phase, etc. 
These parameters and the simulated images are presented as follows.

\subsection{Definition of Flash Models}

There flash models defined by a set of typical input values are considered as an example. 
The used parameters are listed in Table~\ref{tab:table2}.

\begin{figure}
\centering

\includegraphics[width=0.5\textwidth]{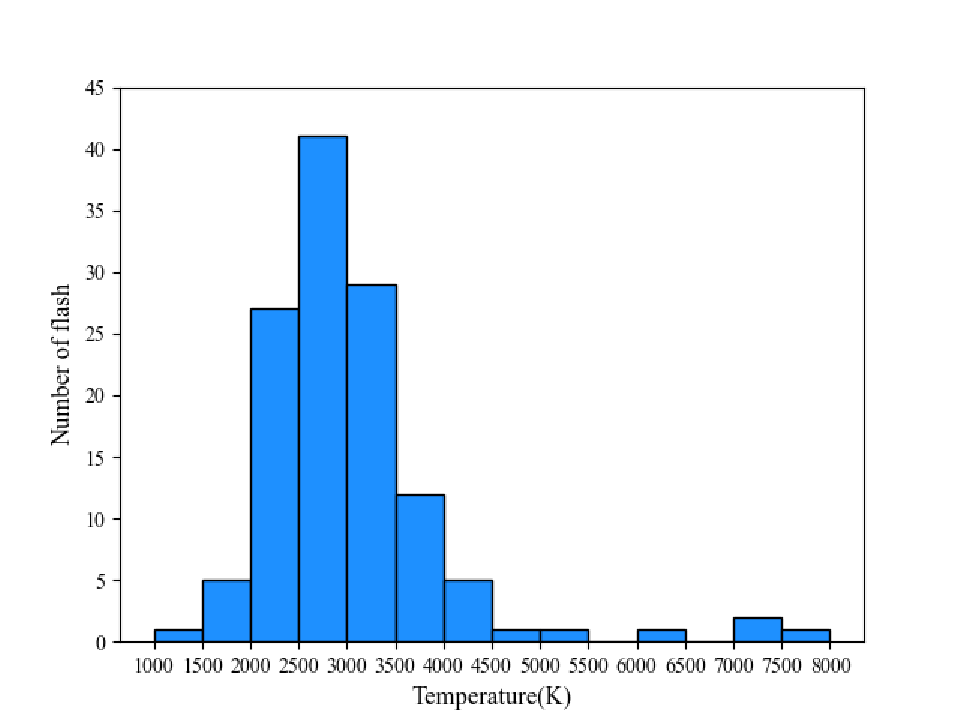}
\caption{Distribution of the peak temperature of the 126 flashes observed on the ground by NELIOTA, after excluding the 61 events with large relative uncertainties of the estimated temperature (i.e., $\geq20\%$). 
We refer the readers to Appendix A for the details of the temperature estimation.}
\label{fig3}
\end{figure}

In the droplet model given in \citep{cintala1992impact}, the temperature ranges from 1700~K (the melting temperature of the lunar regolith) to 3800~K (the evaporation temperature). This range 
agrees with our statistical study on the 187 events recorded by NELIOTA. After excluding 61 
events with large relative uncertainties (i.e., $\geq20\%$), our study shows that 
$\sim85\%$ impacts have a peak temperature between 1700 and 3800~K, and the average value is $\sim2750$~K (see \autoref{fig3}). 

Regarding the meteoroid mass, \cite{liakos2020neliota} estimates a rather wide range
between 2.3~g and 2.7~kg from validated flash data, although the majority ($\sim$71\%) has a mass less than 100~g.

The impact velocity of meteoroids depends on their origin, which is set between 
$15\ \mathrm{km\ s^{-1}}$ for sporadic events \citep{drolshagen2020velocity} and $\sim70\ \mathrm{km\ s^{-1}}$ for the Leonid meteor shower.
An average velocity of $46.3\ \mathrm{km\ s^{-1}}$ can be obtained from a sample of 55 meteoroids
whose velocities are estimated by \cite{avdellidou2019temperatures}.

The radius of individual small droplets produced by an impact is believed to be compared to 
the average grain size of lunar soil, which is approximately tens to hundreds of micrometers \citep{mckay1991lunar}. For the density of meteoroids, we adopt the widely used assumption that meteoroids are composed of diabase with a density of $3.0\ \mathrm{g\ cm^{-3}}$ 
\citep{cintala1992impact}, although the density is approximately $1.0\ \mathrm{g\ cm^{-3}}$ for common comet and is
up to $6.0\ \mathrm{g\ cm^{-3}}$ for iron meteoroids.

\begin{table}[htbp]
  \centering
  \caption{The three flash models considered in the simulation.}
  \label{tab:table2}
  \begin{tabular}{cccccc}
    \toprule
    Flash & $T_0$ (K) & $m$ (g) & $\upsilon$ (km/s) & $V$ (m$^3$) & $r$ ($\mu$m) \\
    \midrule
    1 & 1700 & 2.3 & 15 & 0.000014 & 100 \\
    2 & 2750 & 28 & 46.3 & 0.0019 & 80 \\
    3 & 3800 & 2700 & 70 & 0.2 & 50 \\
    \bottomrule
  \end{tabular}
\end{table}

\subsection{Definition of Observation Equipment}

The parameters adopted to define an observation equipment is tabulated in Table~\ref{tab:table3}. 
The corresponding field-of-view (FoV) is \( 1.47^\circ \times 1.47^\circ \). The angular diameter of the Moon as seen from the L2 point is about \( 3^\circ \), so the FoV covers about 1/4 of the Moon's surface.
This FoV enables us to monitor the non-illuminated part of the Moon efficiently. 

In addition, both Johnson-Cousins $R-$ and $I-$bands are adopted in the subsequent image simulation.

\begin{table}[htbp]
  \centering
  \caption{Optical and camera parameters considered in the simulation.}
  \label{tab:table3}
  \resizebox{\columnwidth}{!}{
    \begin{tabular}{@{}ll|ll@{}}
      \toprule
      \multicolumn{2}{c|}{Optical system parameters} & \multicolumn{2}{c}{Camera parameters} \\ \midrule
      Lens aperture       & 200 mm       & Sensor size    & 7.5 $\mu$m \\
      Focal length        & 600 mm       & Active Pixels  & 2048 $\times$ 2048 \\
      System focal ratio  & $f$/3.0       & Frame frequency & 30 Hz \\
      FOV                 & 1.47$^{\circ}$$\times$ 1.47$^{\circ}$ & Exposure time  & 23 ms \\ 
      Average throughput  & 40\%  & Average quantum efficiency     & 90\% \\ 
      $\lambda_R$ & 641 \text{nm} & Gain   & 1.0 e$^{-}$ per A/D count  \\
      $\lambda_I$ & 798 \text{nm} & Read noise & 6.0 e$^{-}$ rms\\ 
                  &        &   Dark noise    & 0.1 e$^{-}$ pixel$^{-1}$s$^{-1}$  \\ \bottomrule
    \end{tabular}
  }
\end{table}
\subsection{Observation Conditions}

\begin{figure}
\centering
\includegraphics[width=0.5\textwidth]{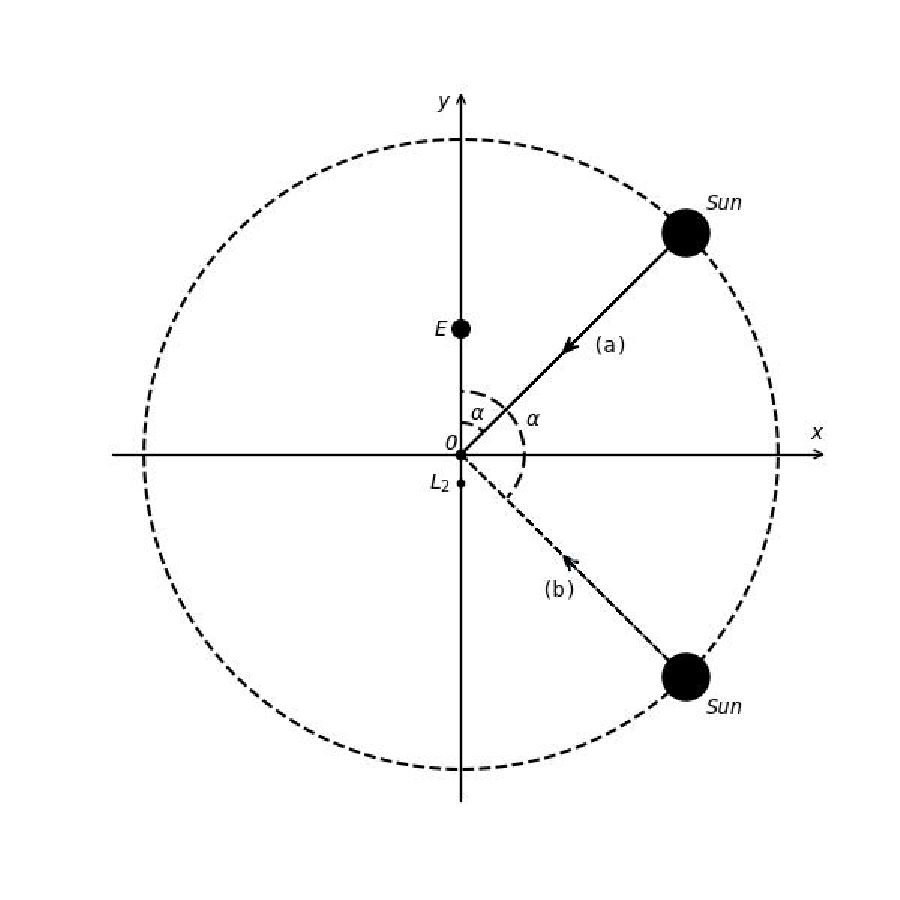}
\caption{An illustration of the geometry adopted in the simulator for stray light estimation. The Moon is at the center of the $xy$ coordinate system.  The Earth and the L2 point are denoted by 
$E$ and $L_2$ on the $y$ axis. The incident radiation from the Sun at two different lunar phases 
is shown by the solid and dashed lines associated with an arrow. The corresponding source 
incident angle $\alpha$ are marked on the plot.
}
\label{figgeometry}
\end{figure}

In the previous section, we introduced the definition of PST. However, determining PST requires specialized simulation software or experimental measurements. Here, we referenced the experimental results of \cite{sholl2007stray}, which demonstrate how PST changes with the angle $\alpha$. Based on these results, along with the geometry shown in \autoref{figgeometry} and the source angle $\alpha$, we estimated the PST values for the Sun, Earth, and Moon at different moon phases. The estimated values we used are presented in Table~\ref{tab:table4}. Although this approach is imperfect, as mentioned in Section 6, we plan to refine the PST model in our future work.

\begin{table}[htbp]
  \centering
  \caption{The PST of the Sun, Earth, and Moon at different moon phases (where 0 = new Moon, 1 = full Moon).}
  \label{tab:table4}
  \begin{tabular}{cccc}
    \toprule
    Moon phase & PST\textsubscript{Sun} & PST\textsubscript{Earth} & PST\textsubscript{Moon} \\
    \midrule
    0.1        & \(2 \times 10^{-5}\) & \(8 \times 10^{-4}\) & \(1 \times 10^{-2}\) \\
    0.5        & \(1 \times 10^{-9}\) & \(5 \times 10^{-3}\) & \(3 \times 10^{-2}\) \\
    \bottomrule
  \end{tabular}
\end{table}

\subsection{Results}

With the input parameters given above and the theoretical model, 
the final images in the 
$R-$ and $I-$bands at 0.1 and 0.5 moon phases are simulated for the Flash 1, 2, and 
3 models listed in Table~\ref{tab:table2}. Table~\ref{tab:flash_snr} tabulates the highest signal-to-noise ratio (SNR) estimated from the simulated images in different cases. The SNR of the simulated image was calculated using the standard definition, as outlined by \cite{raab2002detecting}, where SNR is the signal in $e^-$ divided by the Poisson noise in $e^-$ RMS (Root Mean Square). The flash intensity is measured using standard aperture photometry with a 2-pixel aperture radius, while the background intensity is assessed using annuli ranging from 6 to 14 pixels. This choice reduces contamination from the flash signal to accurately measure background intensity.

\begin{figure}
\centering
\includegraphics[width=0.5\textwidth]{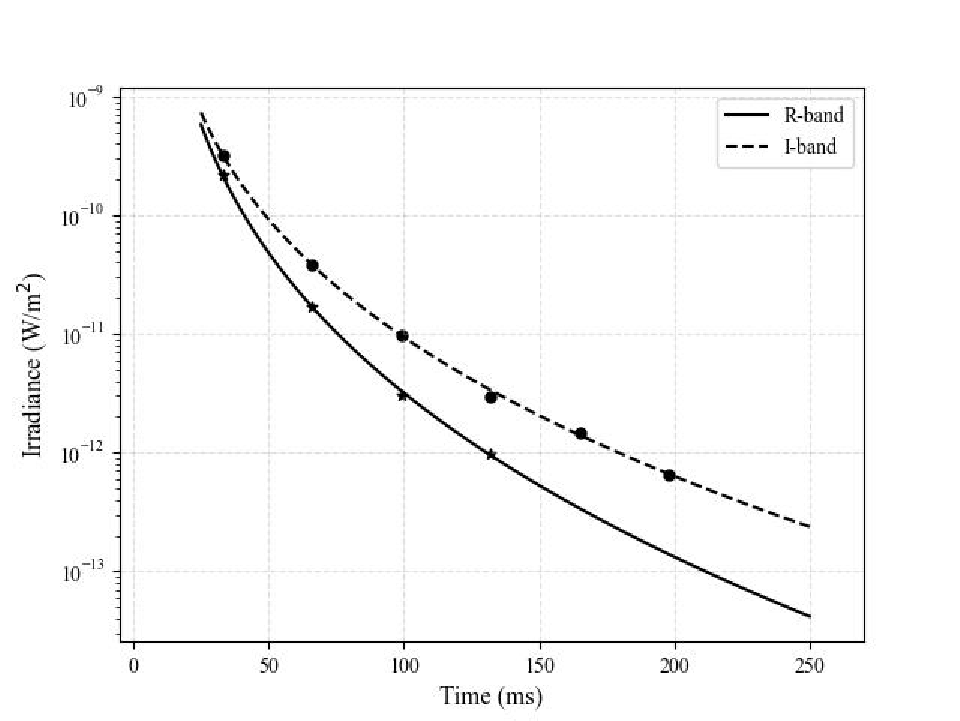}
\caption{The modeled evolution of the irradiance, received at the L2 point, of the Flash 2 model. The solid line represents the $R-$band light curve, and the dashed line the $I-$band. Time = 33 ms marks the first frame, with each subsequent frame spaced 33 ms apart. The overplotted stars and circles
correspond to the simulated images shown in Figures 9 and 10, respectively.}
\label{fig_lightcurve}
\end{figure}

\begin{figure*}
\centering
\includegraphics[width=\textwidth]{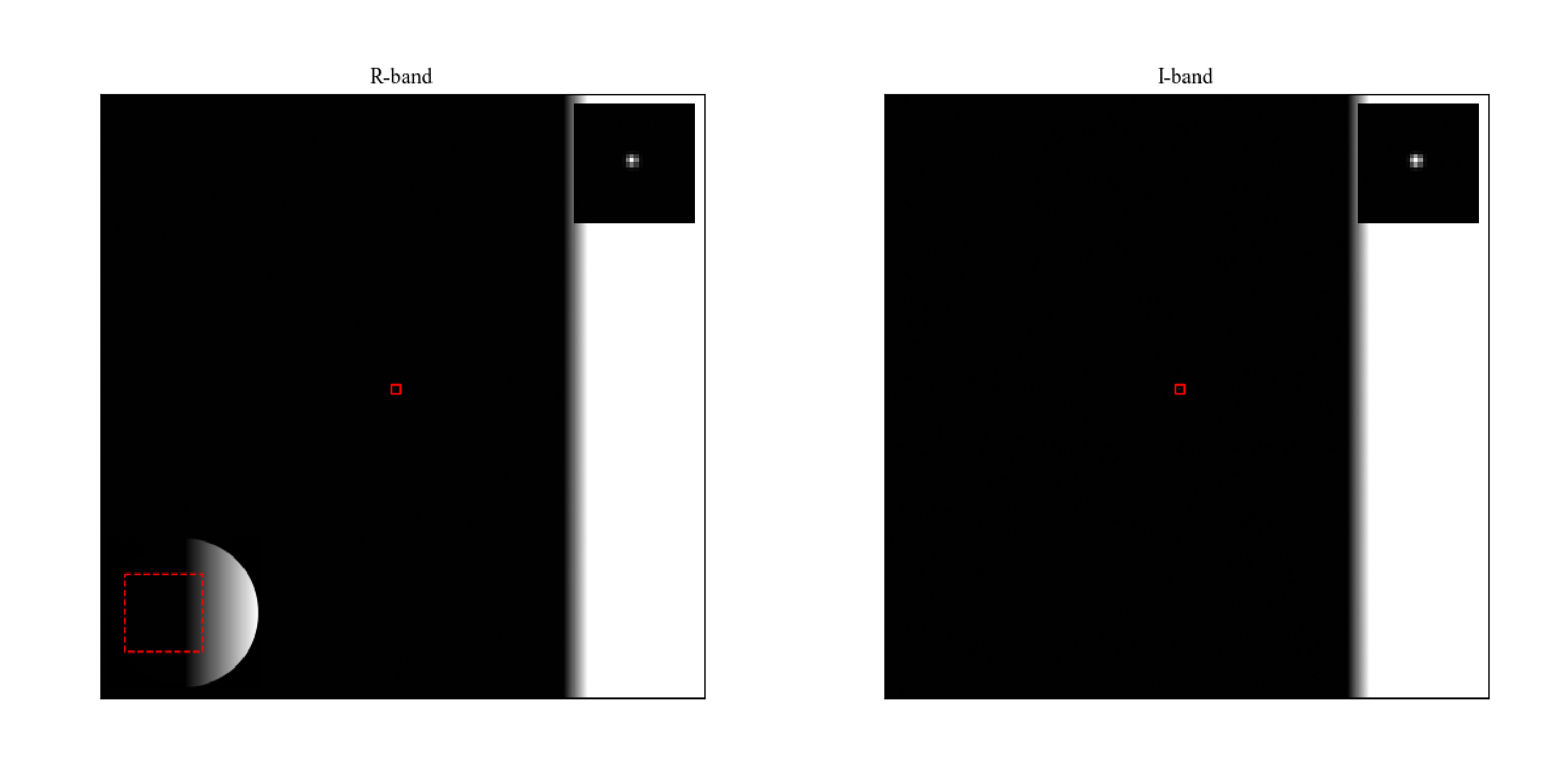}
\caption{Simulating a full-size image of \(2048 \times 2048\) pixels at 0.5 moon phase using the flash 2 model. A \(30 \times 30\) pixel window is used to highlight the flash, with the left side displaying the R-band simulated image and the right side showing the I-band simulated image, and both images are the first frame. The corresponding field of view for the entire image is indicated by the red dashed lines in the lower left corner.}
\label{window}
\end{figure*}

\begin{figure*}[htbp]
  \centering
  \includegraphics[width=\textwidth]{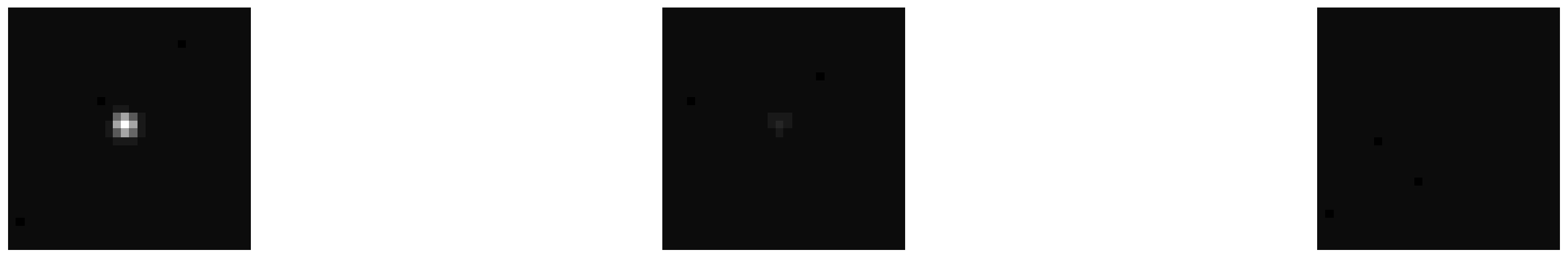}
  \caption{An evolution of the $R-$band images simulated for the flash 2 model occurring at 0.1 moon phase. The time increases from left to right. Only the 30$\times$30 pixels area centered on the flash are displayed in the figure. }
  \label{fig7}
\end{figure*}

\begin{figure*}[htbp]
  \centering
  \includegraphics[width=\textwidth]{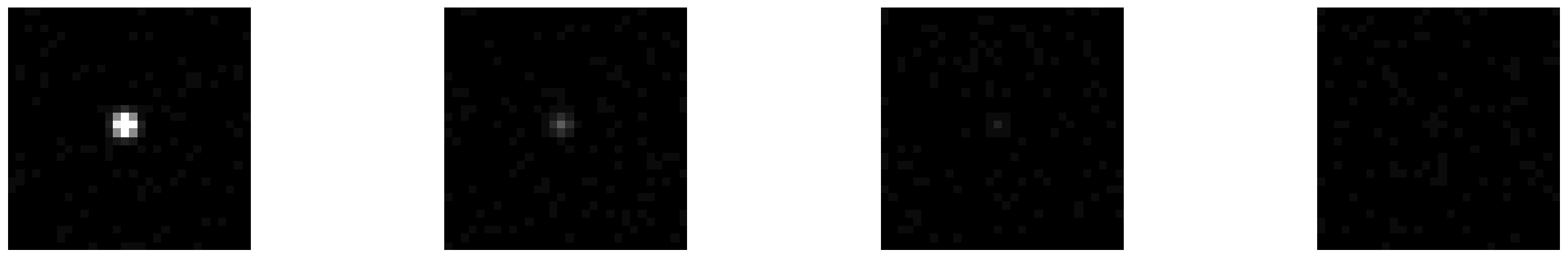}
  \caption{The same as Figure 7, but for the $I-$band.}
  \label{fig8}
\end{figure*}

\begin{figure*}[htbp]
  \centering
  \includegraphics[width=\textwidth]{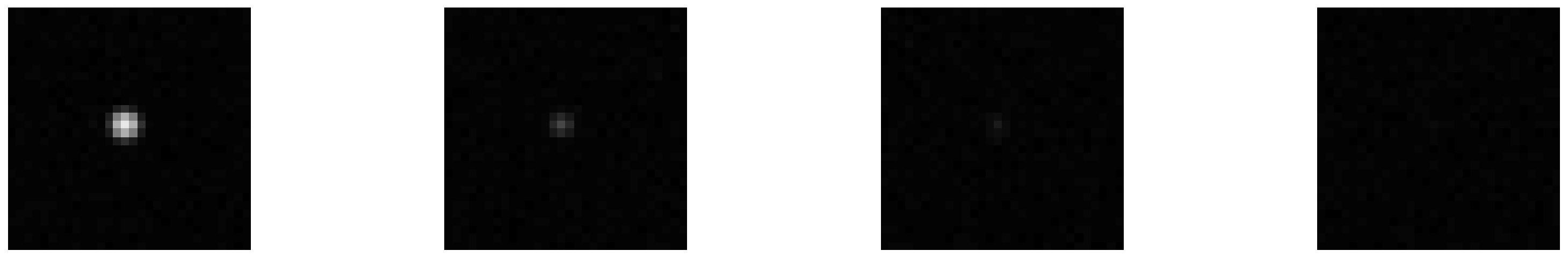}
  \caption{The same as Figure 7, but for a flash occurring at 0.5 moon phase.}
  \label{fig9}
\end{figure*}

\begin{figure*}[htbp]
  \centering
  \includegraphics[width=\textwidth]{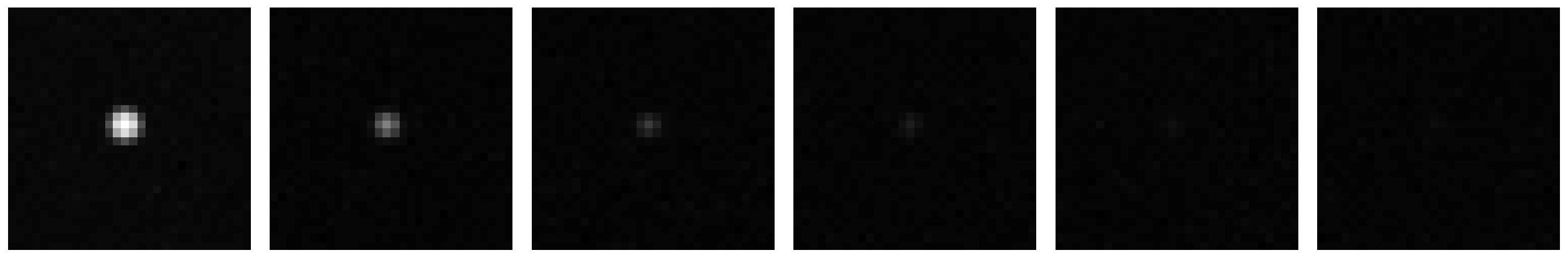}
  \caption{The same as Figure 9, but for the $I-$band. }
  \label{fig10}
\end{figure*}

As examples, \autoref{fig_lightcurve} shows the calculated evolution of irradiance of the Flash 2 model. Then, we simulated a full-size image (\(2048 \times 2048\)) that includes the illuminated area of the lunar surface, as shown in \autoref{window}. It is important to note that during actual observations, the illuminated areas should be avoided, and the field of view should be directed towards the non-illuminated part. The presence of illuminated areas within the field of view significantly reduces the efficiency of monitoring lunar impact flashes. Therefore, the NELIOTA project conducts flash observations in shadow regions when the moon phase ranges from 0.1 to 0.45 \citep{xilouris2018neliota}. \autoref{window} is designed to help readers understand the relative intensity between the brightness of the flash and the illuminated limb of the lunar surface.  The subsequent analysis focuses on simulating the temporal variations of the flash in the shadow region. 
By extracting the 30$\times$30 pixel area centered on the flash, 
\autoref{fig7} and \ref{fig8} illustrate the temporal variations of the simulated Flash 2 images taken at 
the 0.1 moon phase.  Similar images taken at the 0.5 moon phase are displayed in 
\autoref{fig9} and \ref{fig10}. 

Three facts can be learned from the simulated images and the corresponding SNRs. 
At fitst, the flashes generated by meteoroid impacts 
tend to be more easily detected in the $I-$band than the $R-$band, which is consistent with 
the trend observed by NELIOTA project: the flashes detected
in the $I-$band outnumber those detected in the $R-$band. Possible reasons include: 
1) the lunar surface brightness in the $I-$band is lower than that in the R-band, resulting 
in weaker stray light in the $I-$band;  2) the radiation of the flashes is more prominent in 
the near-infrared. 

Secondly, in a given band, the number of the flashes detected at the 0.5 moon phase is slightly higher than that at the 0.1 moon phase, which is mainly caused by a reduced stray light (i.e.,
a reduced PST value) at the 0.5 moon phase.   

Finally, an event described by the Flash 1 model is hard to
 detect by the instrument adopted in the current simulation at the L2 point. In fact, 
our further simulations show that a telescope with a minimum aperture of 537 millimeters 
is required to capture the faint event with a SNR threshold of 3.0.  

When the temperature exceeds 3800~K, the droplet model described in Section 3.1 is 
no longer valid because vapor is believed to be generated upon impact \citep{cintala1992impact}. 
This case can be 
covered easily by our simulator by slight modifications. Specifically speaking, 
the total volume of the vapor produced upon impact can be calculated by Eq. (8), but with 
different constants provided in \cite{cintala1992impact}. The radius and the area of the sphere can then be easily determined by assuming the vapor has a regular spherical shape. 

\begin{figure*}[htbp]
  \centering
  \includegraphics[width=\textwidth]{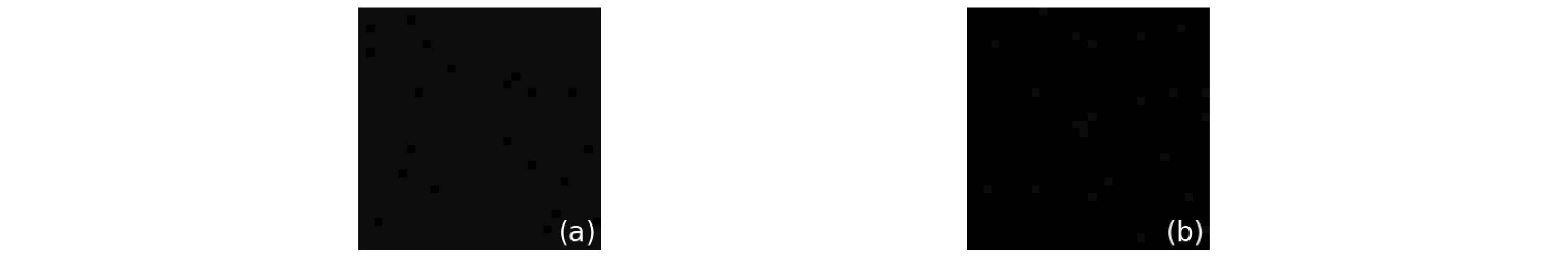}
  \caption{Simulated images taken in $R-$ (panel a) and $I-$bands (panel b) of a flash 
  with $T_0=4540$K. All the images are simulated at 0.1 moon phase. Due to the SNR being just over 3.0, the flash is difficult to discern by the naked eye, but it can be easily identified using software.}
  \label{fig4540k}
\end{figure*}

As an example, \autoref{fig4540k} illustrates the simulated flash images in both $R-$ and 
$I-$bands for a flash with $T_0=4540$~K\footnote[4]{The flash No. 43 in \cite{avdellidou2019temperatures}.}, for which the mass and velocity of the flash are 13.43 g and 58.35 km/s, respectively.
In the current simulation, $m_\rho$ and $c_\mathrm{p}$ are adopted to be 
$0.2 \, \text{g/cm}^3$ and $0.67 \, \text{J} \cdot \text{g}^{-1} \cdot \text{K}^{-1}$,
respectively, for gaseous silicon dioxide.  The corresponding values of SNR are 
measured to be 3.40 and 3.81 for the $R-$ and $I-$bands, respectively, which are much lower
than the case of the Flash model 2 (Table~\ref{tab:flash_snr}). The low SNR mainly results from a significant decrease of surface area in the vapor model.


\begin{table}[htbp]
\centering
\caption{Values of SNR determined from the simulated images for flash at different moon phases and bands.}
\begin{tabular}{cccccc}
\toprule
\multirow{2}{*}{Flash} & \multicolumn{2}{c}{0.1 moon phase} & \multicolumn{2}{c}{0.5 moon phase} \\ \cmidrule(lr){2-3} \cmidrule(lr){4-5}
                        & $R-$band                        &       $I-$band                 &  $R-$band                      &  $I-$band                       \\ \midrule
1                       & 0.48       & 0.66                          & 1.02                          & 1.32                          \\
2                       & 228                            & 360                           & 383                           & 525                           \\
3                       & 525                          & 567                         & 888                         & 891                         \\ \bottomrule
\end{tabular}
\label{tab:flash_snr}
\end{table}

\section{Validation}

We validate our simulator in this section by focusing on the three aspects:  the background statistics, the simulated PSF and a comparison with 
the flashes captured on the ground.
\subsection{Background statistics}
\begin{figure}[htbp]
  \centering
  \includegraphics[scale=0.35]{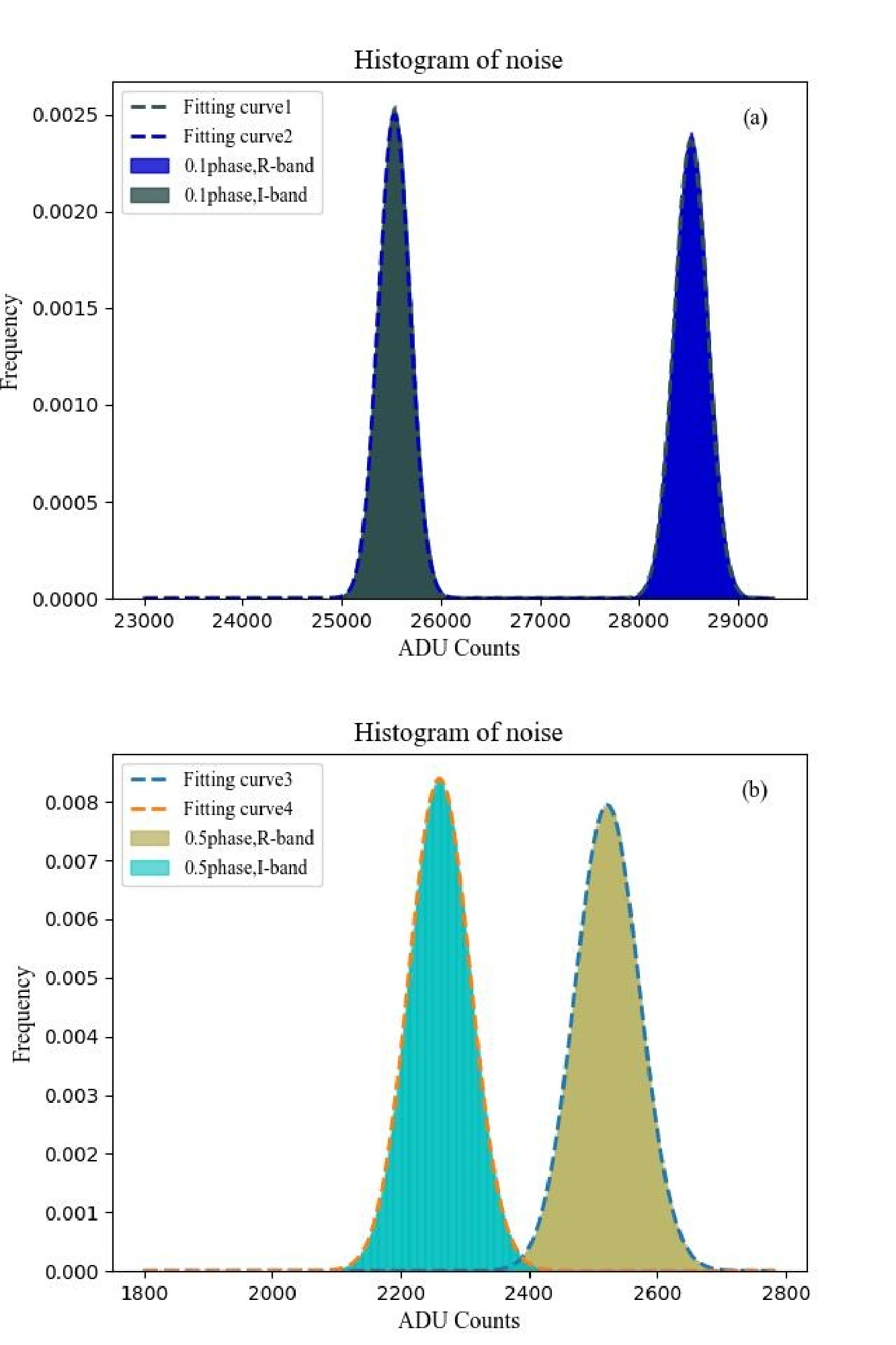}
  \caption{\it Panel (a): \rm Distribution of the $R-$ and $I-$bands background levels simulated at the 0.1 moon phase. The best-fit Poisson distributions are over-plotted by the dashed lines; \it Panel (b): \rm the same as the Panel (a), but for the images simulated at moon phase of 0.5.} 
  \label{figbackground}
\end{figure}

\autoref{figbackground} (a) and (b) show that distributions of the simulated background at the 0.1 and 0.5 moon phases, respectively, along with the best-fit Poisson distributions. 
By using the least squares method, the best-fit mean values in ADU are 28521.95, 25529.96, 2522.21, and 2261.09, which are highly consistent with the input values of 28521.86, 25529.95, 2522.20, and 2261.09. The corresponding residual mean squares are calculated to be $2.6\times 10^{-7}$, $2.3\times 10^{-7}$, $2.4\times 10^{-7}$, and $2.1\times 10^{-7}$.

\subsection{PSF profile of the flash}

\begin{figure*}[htbp]
  \centering
  \includegraphics[width=\textwidth]{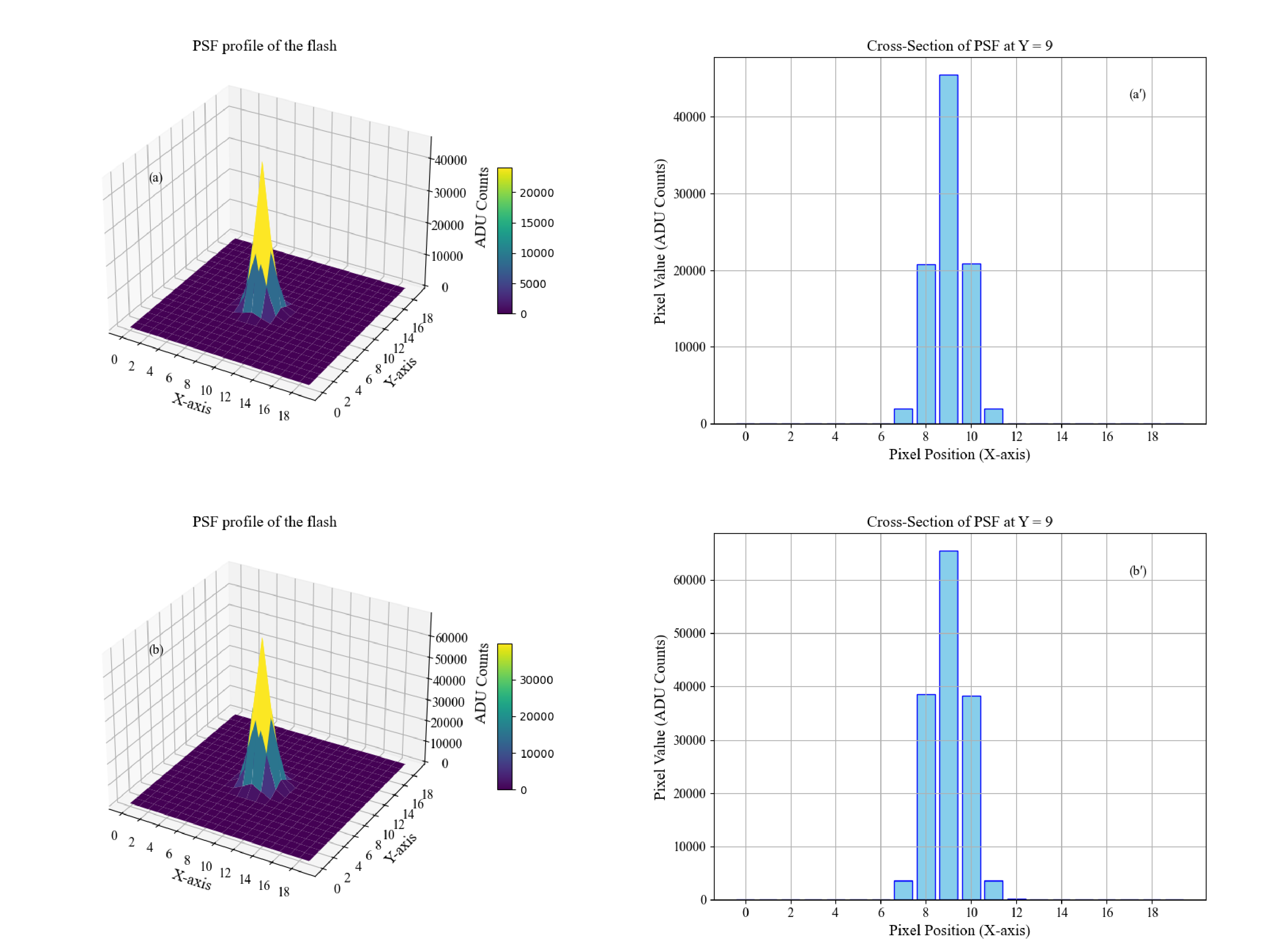}
  \caption{The simulated PSF in $R-$ (panel a) and $I-$bands (panel b) of the first frame images of the flash 2 model. Additionally, we plotted histograms to display the ADU counts of individual pixels. Panel a$^{\prime}$ and panel b$^{\prime}$ present the histograms for the R-band and I-band, respectively, both of which are cross-sections taken at y = 9 (where the coordinates of the flash center are (9, 9)).}
  \label{figPSF}
\end{figure*}

The simulated PSF profiles of the flash 2 model are illustrated in \autoref{figPSF}. After fitting each simulated PSF by a two-dimensional Gaussian function, Table~\ref{tab:table6} presents a comparison between the fitting parameters and the inputs, which again indicates a high consistency between the simulated and input PSF. 

\begin{table}[h]
\centering
\caption{Comparison of the fitted and input PSF in $R-$ and $I-$bands.}
\label{tab:table6}
\renewcommand{\arraystretch}{1.0}
\begin{threeparttable}
\begin{tabular}{cccccc}
\hline
Parameters$^a$ & $G_0$    & $x_0$   & $y_0$   & $\sigma_x$ & $\sigma_y$ \\ \hline
Input(R)   & 45202 & 9.0    & 9.0    & 0.80     & 0.80     \\
Fitted(R) & 45515 & 9.002    & 8.996    & 0.7996   & 0.7938   \\
Input(I)   & 65535 & 9.0    & 9.0    & 0.80     & 0.80     \\
Fitted(I) & 69031 & 8.998 & 9.003 & 0.8744   & 0.8750   \\ \hline
\end{tabular}
\begin{tablenotes}
 \item[a] "Input (R)" or "Input (I)" refers to the input parameters for the R-band or I-band, respectively. These parameters are \(G_0\), \(x_0\), \(y_0\), \(\sigma_x\), and \(\sigma_y\). Specifically, \(G_0\) is the maximum value at the center of the flash, \(x_0\) and \(y_0\) represent the coordinates of the flash center, and \(\sigma_x\) and \(\sigma_y\) are the dispersion in the x and y directions, respectively. "Fitted (R)" or "Fitted (I)" represents the parameters obtained from fitting.
\end{tablenotes}
\end{threeparttable}
\end{table}

\subsection{Comparison with ground observations}

In order to validate our simulator further, we run the simulator with a series of input models, and compare the results obtained from the simulations with the
values measured from the observations. Three flashes (Flash A, D and E)
modeled in detail by \cite{yanagisawa2002lightcurves} are adopted in our comparison study.
With the modeled parameters listed in Table~\ref{tab:table7}, the total energy per unit area received on the Earth is at first
calculated to be $1.45 \times 10^{-12}$, $8.51 \times 10^{-12}$ and $1.9 \times 10^{-11}\ \mathrm{J\ m^{-2}}$ for the Flash A, D and E, respectively, which agrees well with the observed values (See Table~\ref{tab:table7}).

\begin{table*}[h]
\caption{Summary of flash parameters, observed and simulated energy per unit area.}
\label{tab:table7}
\centering
\renewcommand{\arraystretch}{1.2}
\begin{threeparttable}
\begin{tabular}{cccccc}
\hline
Flash & $T_0$ (K) & $R_d$ ($\mu$m) & $V$ (m$^3$) & $\Phi_{\text{obs}}$ (J/m$^2$) & $\Phi_{\text{sim}}$ (J/m$^2$) \\
\hline
A     & 2100      & 100            & 0.05        & $2.0 \times 10^{-12}$ & $1.45 \times 10^{-12}$ \\
D     & 2100      & 150            & 0.3         & $> 7.3 \times 10^{-12}$ & $8.51 \times 10^{-12}$ \\
E     & 2250      & 180            & 0.4         & $> 1.1 \times 10^{-11}$ & $1.90 \times 10^{-11}$ \\
\hline
\end{tabular}
\begin{tablenotes}
 \item $\Phi_{\text{obs}}$ represents the observed energy per unit area, while $\Phi_{\text{sim}}$ represents the simulated.
\end{tablenotes}
\end{threeparttable}
\end{table*}

We subsequently simulate the images of the flashes A, D, and E using parameters similar to those of the observation system \uppercase\expandafter{\romannumeral1} used in \cite{yanagisawa2002lightcurves}, and calculate the corresponding apparent magnitudes by using Vega as the standard star. 
These calculated magnitudes are compared with the measured brightness provided by \cite{yanagisawa2002lightcurves} in Table~\ref{tab:table8}. The comparison shows a rough match 
between the simulated and observed values, which is acceptable because the extinction due to earth
atmosphere is not included in our simulator.
\begin{table}[h]
\caption{Comparison of flash brightness measured from the simulated and observed images.}
\label{tab:table8}
\centering
\begin{threeparttable}
\renewcommand{\arraystretch}{1.5}
\resizebox{\linewidth}{!}{
\begin{tabular}{cccccccc}
\hline
Flash$_a$               & $A_1$   & $D_1$ & $D_2$ & $E_1$ & $E_2$ & $E_3$ & $E_4$  \\ \hline
Simulated magnitude & 6.32    & 4.82  & 5.33  & 4.05  & 4.54  & 4.99  & 5.41  \\
Observed magnitude  & 6       & \multicolumn{6}{c}{$\leq 5$} \\ \hline
\end{tabular}}
\begin{tablenotes}
  \item[a] The subscript $i$ denotes the observed $i$th image. 
\end{tablenotes}
\end{threeparttable}
\end{table}

\section{Conclusion and Prospect}

In this study, we present an end-to-end image simulator dedicated to capturing impact flashes on lunar far side from space, which is useful for future mission 
design and optimization. 
The simulator is designed using a modular approach, developed in Python environment and 
validated by comparing the simulated results with ground-based observations.
The simulator encompasses modules for flash model, background radiation, telescope and detector model. 

The modular design enables us to easily improve and update the simulator in the future by adding or replacing more complex physical models in the following aspects:
\begin{itemize}
   \item Photon counts prediction with spectral energy distributions of both flash and background emission, more detailed and specific flash model and specific transmission of the system.
   \item More realistic model for the reflection of sunlight by the lunar surface. Rather than the simple Lambert model and the average albedo, the real reflection of lunar surface is complex and depends on lunar topography, morphology, and wavelength. 
   \item Sky background emission. An interface has been designed to incorporate the sky 
   background in the future, which can create a more complete image that includes the area around the Moon and can also be used for simulating ground-based telescopes, although its level is believed to be much lower than that of lunar surface reflection. 
   \item Imperfect optical system including vignetting, PSF distortion and PST curve 
   either modeled or measured. 
   \item  Detector pixel imperfections (Photo Response Non-Uniformity and Dark Signal Non-Uniformity) and dependence of 
   dark current and gain on temperature. 
   \item  CCD smear simulation. This simulation models the smear effect introduced by the line-by-line readout process of CCD sensors. During this process, the illuminated part of the Moon may leave residual traces on other lines, particularly under high brightness, degrading image quality. This effect, due to sequential line readout, may not occur with CMOS sensors, which employ a simultaneous frame readout method.
\end{itemize}

\section*{Acknowledgments}
This study is supported by the Natural Science Foundation under grant No. 11973063
and by the Strategic Pioneer Program on Space Science, Chinese
Academy of Sciences, grants XDA15072119.
JW is supported by Natural Science Foundation of Guangxi (2020GXNSFDA238018). We also sincerely thank the reviewers for their valuable insights and suggestions, which have greatly contributed to improving the quality of this paper.

\appendix
\section{Tempratures of the Flashes Observed by NELIOTA}

The NELIOTA project, funded by the European Space Agency, has monitored lunar impact flashes in standard Johnson-Cousins $R-$ and 
$I-$band filters using the 1.2 m Kryoneri telescope located at the National Observatory of Athens, 
Greece, since 2015 \citep{bonanos2018neliota,xilouris2018neliota,liakos2019neliota}. The primary performance of the project are outlined in Table~\ref{tab:tableA9}. 
Its large FoV of \( 17.0' \times 14.4' \) enables NELIOTA to monitor up to one-third of the 
lunar nearside surface. 
Up to July 2023, a total of 187 events has been detected in both bands within 
approximately 278 hours observation
, which leads to an overall rate of verified flash events of approximately one flash per 1.5 hours.

By assuming a blackbody emission, 
the temperature of each of the verified flashes reported by NELIOTA can be estimated by us from 
the flux ratio \citep{liakos2020neliota} of the two bands: 
\begin{equation}
\centering
\frac{f_R}{f_I} = \left(\frac{\lambda_I}{\lambda_R}\right)^5 \cdot \frac{\mathrm{e}^{\frac{hc}{\lambda_I k_\mathrm{B} T}} - 1}{\mathrm{e}^{\frac{hc}{\lambda_R k_\mathrm{B} T}} - 1}
\label{eq18}
\end{equation}
where $\lambda_R$ and $\lambda_I$ are the effective wavelengths of the two used filters.
With the zero-points of the two bands \citep{bessell1998model}, the flux can be obtained from the corresponding magnitudes, $m_R$ and $m_I$:
\begin{equation}
\centering
f_R = 2.18 \times 10^{-8} \times 10^{-0.4m_R} \, \mathrm{in\ W\ m^{-2}\mu m^{-1}}
\label{eq15}
\end{equation}
\begin{equation}
\centering
f_I = 1.13 \times 10^{-8} \times 10^{-0.4m_I} \, \mathrm{in\ W\ m^{-2}\mu m^{-1}}
\label{eq16}
\end{equation}
We calculate the uncertainty of the estimated temperature by a Monte Carlo simulation with 
$10^{5}$ iterations. In each iteration, a new set of $m_R$ and $m_I$ is derived by a random sampling, which is based on the Gaussian distributions defined by the photometric errors 
in both bands provided by NELIOTA.

The estimated peak temperatures, along with the corresponding uncertainties at $1\sigma$ 
significance level, are listed in Table~\ref{tab:tableA10}, where we also include flash temperatures calculated by \cite{liakos2020neliota,liakos2024neliota}. Note that in calculating the flash temperatures, we used the constants from the SciPy library, specifically \( h = 6.62607015 \times 10^{-34} \, \mathrm{J} \cdot \mathrm{s} \), \( c = 299792458 \, \mathrm{m \cdot s^{-1}} \), and \( k_B = 1.380649 \times 10^{-23} \, \mathrm{J \cdot K^{-1}} \). By contrast, \cite{liakos2020neliota,liakos2024neliota} used approximate values: \( h = 6.63 \times 10^{-34} \, \mathrm{J} \cdot \mathrm{s} \), \( c = 2.998 \times 10^8 \, \mathrm{m \cdot s^{-1}} \), and \( k_B = 1.38 \times 10^{-23} \, \mathrm{J \cdot K^{-1}} \). This difference, however, is negligible compared to the original temperature uncertainties.

For most temperatures (182 out of 187), only minor discrepancies are observed. However, we identified five flashes with notable temperature differences: IDs 17, 83, 108, 147, and 181. For ID 17, we found that the flash magnitudes reported by \cite{liakos2020neliota} in Table A.1 ($m_R$ = 10.92 and $m_I$ = 9.20) differ from those on the NELIOTA website ($m_R$ = 10.46 and $m_I$ = 9.03), and we used the NELIOTA values. ID 83 corresponds to flash ID 118 in Tables A.1 and A.2 of \cite{liakos2024neliota}. Table A.1 shows differences in magnitudes and magnitude errors for flash IDs 118 and 120, while Table A.2 provides identical calculated values for flash temperature and temperature error for both flashes, which likely indicates an error.

For the remaining flashes with significant temperature differences (IDs 108, 147, and 181), we have not yet identified the cause, but we have conducted thorough checks to ensure the accuracy of our calculations.

\begin{table}[htbp]
  \centering
  \caption{The main technical specifications of the NELIOTA telescope.}
  \label{tab:tableA9}
  \resizebox{\columnwidth}{!}{%
    \begin{tabular}{@{}ll|ll@{}}
      \toprule
      \multicolumn{2}{c|}{Optical system parameters} & \multicolumn{2}{c}{Camera parameters} \\ \midrule
      Primary mirror     & 1200 mm      & Sensor type    & Front-illuminated Scientific CMOS \\
      System focal ratio & $f$/2.8       & Sensor size    & 6.48 m \\
      FoV                & 17.0' $\times$ 14.4'  & Shutter        & Global \\
                         &             & Valid Pixels   & 1280 $\times$ 1080 \\
                         &             & Frame frequency & 30 Hz \\
                         &             & Exposure time   & 23 ms \\ \bottomrule
    \end{tabular}%
  }
\end{table}

\begin{table*}[htbp]
\centering
\caption{As of June 22, 2023, NELIOTA has recorded a total of 187 flash events. This table presents the peak magnitudes and occurrence times for each flash as recorded, along with the flash temperatures and their corresponding temperature errors, which were calculated separately by us and \cite{liakos2020neliota,liakos2024neliota}.}
\label{tab:tableA10}
\begin{tabular}{@{}lcccccc@{}}
\toprule
Flash(ID) & Date(UT) & Time(UT) & R $\pm\ \sigma_R$(mag) & I $\pm\ \sigma_I$(mag) & T $\pm\ \sigma_T$(K) & T $\pm\ \sigma_T$(K) \citep{liakos2020neliota,liakos2024neliota} \\
\midrule
1 & 2017/2/1 & 17:13:58 & 10.15$\pm$0.12 & 9.05$\pm$0.05 & 3047$\pm$263 & 3046$\pm$307 \\
2 & 2017/3/1 & 17:08:47 & 6.67$\pm$0.07 & 6.07$\pm$0.06 & 4509$\pm$427 & 4503$\pm$646 \\
3 & 2017/3/1 & 17:13:17 & 9.15$\pm$0.11 & 8.23$\pm$0.07 & 3445$\pm$341 & 3438$\pm$431 \\
4 & 2017/3/4 & 20:51:32 & 9.5$\pm$0.14 & 8.79$\pm$0.06 & 4073$\pm$598 & 4076$\pm$748 \\
5 & 2017/4/1 & 19:45:52 & 10.18$\pm$0.13 & 8.61$\pm$0.03 & 2344$\pm$156 & 2343$\pm$200 \\
6 & 2017/5/1 & 20:30:58 & 10.19$\pm$0.18 & 8.84$\pm$0.05 & 2627$\pm$283 & 2615$\pm$315 \\
7 & 2017/6/27 & 18:58:27 & 11.07$\pm$0.32 & 9.27$\pm$0.06 & 2107$\pm$330 & 2101$\pm$343 \\
8 & 2017/6/28 & 18:45:26 & 10.56$\pm$0.38 & 9.48$\pm$0.13 & 3086$\pm$1059 & 3088$\pm$945 \\
9 & 2017/7/19 & 2:00:36 & 11.23$\pm$0.4 & 9.33$\pm$0.06 & 2018$\pm$397 & 2008$\pm$367 \\
10 & 2017/7/28 & 18:21:45 & 11.24$\pm$0.34 & 9.29$\pm$0.06 & 1977$\pm$309 & 1978$\pm$300 \\
11 & 2017/7/28 & 18:42:58 & 10.72$\pm$0.24 & 9.63$\pm$0.1 & 3067$\pm$591 & 3056$\pm$647 \\
12 & 2017/7/28 & 18:51:42 & 10.84$\pm$0.24 & 9.81$\pm$0.09 & 3190$\pm$644 & 3168$\pm$692 \\
13 & 2017/7/28 & 19:17:18 & 8.27$\pm$0.04 & 6.32$\pm$0.01 & 1977$\pm$34 & 1972$\pm$105 \\
14 & 2017/8/16 & 1:05:47 & 10.15$\pm$0.18 & 9.54$\pm$0.1 & 4466$\pm$1048 & 4455$\pm$1146 \\
15 & 2017/8/16 & 2:15:59 & 10.69$\pm$0.28 & 9.11$\pm$0.06 & 2333$\pm$357 & 2326$\pm$356 \\
16 & 2017/8/16 & 2:41:15 & 10.81$\pm$0.3 & 9.08$\pm$0.06 & 2174$\pm$330 & 2167$\pm$319 \\
17 & 2017/8/18 & 2:02:21 & 10.46$\pm$0.02 & 9.03$\pm$0.02 & 2517$\pm$37 & 2185$\pm$255 \\
18 & 2017/8/18 & 2:03:08 & 10.19$\pm$0.12 & 8.83$\pm$0.04 & 2613$\pm$184 & 2615$\pm$256 \\
19 & 2017/9/14 & 3:17:50 & 9.17$\pm$0.07 & 8.07$\pm$0.03 & 3047$\pm$150 & 3058$\pm$207 \\
20 & 2017/9/16 & 2:26:25 & 8.52$\pm$0.03 & 7.04$\pm$0.01 & 2452$\pm$40 & 2440$\pm$136 \\
21 & 2017/10/13 & 1:54:22 & 9.28$\pm$0.11 & 8.37$\pm$0.04 & 3471$\pm$309 & 3458$\pm$357 \\
22 & 2017/10/13 & 2:33:44 & 10.31$\pm$0.24 & 9.89$\pm$0.12 & 5494$\pm$1534 & 5453$\pm$1740 \\
23 & 2017/10/16 & 2:46:46 & 10.72$\pm$0.16 & 9.46$\pm$0.05 & 2764$\pm$280 & 2751$\pm$384 \\
24 & 2017/10/26 & 17:59:43 & 10.03$\pm$0.25 & 9.42$\pm$0.12 & 4466$\pm$1341 & 4431$\pm$1088 \\
25 & 2017/11/14 & 3:34:15 & 10.31$\pm$0.17 & 9.31$\pm$0.06 & 3256$\pm$435 & 3264$\pm$488 \\
26 & 2017/11/23 & 16:17:33 & 10.45$\pm$0.23 & 10.06$\pm$0.12 & 5706$\pm$1541 & 5722$\pm$1528 \\
27 & 2017/12/12 & 2:48:08 & 10.5$\pm$0.24 & 8.98$\pm$0.08 & 2403$\pm$327 & 2402$\pm$334 \\
28 & 2017/12/12 & 4:30:01 & 10.58$\pm$0.28 & 9.84$\pm$0.11 & 3969$\pm$1225 & 3948$\pm$1267 \\
29 & 2017/12/13 & 4:26:58 & 10.56$\pm$0.23 & 9.95$\pm$0.11 & 4466$\pm$1261 & 4432$\pm$1475 \\
30 & 2017/12/14 & 4:35:10 & 7.94$\pm$0.05 & 6.76$\pm$0.02 & 2899$\pm$95 & 2889$\pm$200 \\
31 & 2018/1/12 & 3:54:03 & 10.01$\pm$0.14 & 9.31$\pm$0.07 & 4109$\pm$630 & 4101$\pm$782 \\
32 & 2018/3/23 & 17:24:19 & 9.93$\pm$0.26 & 8.62$\pm$0.06 & 2686$\pm$450 & 2675$\pm$431 \\
33 & 2018/4/10 & 3:36:58 & 8.84$\pm$0.13 & 8.08$\pm$0.05 & 3903$\pm$486 & 3905$\pm$626 \\
34 & 2018/6/9 & 2:29:18 & 9.92$\pm$0.23 & 9$\pm$0.09 & 3445$\pm$740 & 3428$\pm$727 \\
35 & 2018/6/19 & 19:12:10 & 9.87$\pm$0.21 & 9.03$\pm$0.09 & 3659$\pm$766 & 3646$\pm$736 \\
36 & 2018/6/19 & 20:00:48 & 9.92$\pm$0.28 & 9.31$\pm$0.14 & 4466$\pm$1444 & 4436$\pm$1094 \\
37 & 2018/6/19 & 20:04:10 & 10.26$\pm$0.61 & 8.63$\pm$0.11 & 2277$\pm$980 & 2267$\pm$695 \\
38 & 2018/7/9 & 1:44:19 & 11.16$\pm$0.28 & 10.06$\pm$0.12 & 3047$\pm$722 & 3032$\pm$748 \\
39 & 2018/8/6 & 1:57:44 & 9.68$\pm$0.16 & 8.14$\pm$0.04 & 2379$\pm$200 & 2369$\pm$215 \\
40 & 2018/8/6 & 2:38:14 & 9.16$\pm$0.09 & 7.73$\pm$0.02 & 2517$\pm$123 & 2515$\pm$156 \\
41 & 2018/8/7 & 1:33:55 & 10.79$\pm$0.26 & 9.31$\pm$0.07 & 2452$\pm$370 & 2444$\pm$405 \\
42 & 2018/8/7 & 1:35:45 & 8.78$\pm$0.05 & 7.74$\pm$0.02 & 3169$\pm$115 & 3153$\pm$262 \\
43 & 2018/8/7 & 2:33:18 & 10.07$\pm$0.17 & 9.46$\pm$0.07 & 4466$\pm$920 & 4460$\pm$988 \\
44 & 2018/8/7 & 3:10:33 & 10.39$\pm$0.31 & 9.8$\pm$0.14 & 4554$\pm$1527 & 4558$\pm$1087 \\
45 & 2018/8/8 & 2:19:55 & 11.14$\pm$0.28 & 9.9$\pm$0.07 & 2797$\pm$548 & 2793$\pm$583 \\
46 & 2018/8/8 & 2:28:23 & 11.06$\pm$0.21 & 10.4$\pm$0.13 & 4260$\pm$1147 & 4253$\pm$1364 \\
47 & 2018/8/8 & 2:29:45 & 8.36$\pm$0.04 & 7.3$\pm$0.02 & 3127$\pm$93 & 3124$\pm$247 \\
\bottomrule
\end{tabular}
\end{table*}

\clearpage %
\begin{table*}[htbp]
\centering
\begin{tabular}{@{}lcccccc@{}}
\toprule
Flash(ID) & Date(UT) & Time(UT) & R $\pm\ \sigma_R$(mag) & I $\pm\ \sigma_I$(mag) & T $\pm\ \sigma_T$(K) & T $\pm\ \sigma_T$(K) \citep{liakos2020neliota,liakos2024neliota} \\
\midrule
48 & 2018/8/8 & 2:52:26 & 11.05$\pm$0.31 & 9.74$\pm$0.1 & 2686$\pm$588 & 2678$\pm$563 \\
49 & 2018/8/15 & 18:08:17 & 11.8$\pm$0.36 & 9.56$\pm$0.09 & 1765$\pm$262 & 1758$\pm$257 \\
50 & 2018/9/4 & 1:33:53 & 9.87$\pm$0.3 & 9.18$\pm$0.1 & 4145$\pm$1346 & 4163$\pm$1490 \\
51 & 2018/9/5 & 1:51:37 & 7.84$\pm$0.07 & 6.6$\pm$0.02 & 2797$\pm$120 & 2793$\pm$194 \\
52 & 2018/9/5 & 2:47:54 & 10.61$\pm$0.37 & 9.09$\pm$0.09 & 2403$\pm$552 & 2401$\pm$501 \\
53 & 2018/9/6 & 2:00:33 & 10.95$\pm$0.3 & 10.33$\pm$0.14 & 4423$\pm$1469 & 4428$\pm$1348 \\
54 & 2018/9/6 & 3:10:04 & 11.18$\pm$0.25 & 9.86$\pm$0.09 & 2671$\pm$443 & 2660$\pm$459 \\
55 & 2018/10/15 & 18:17:49 & 9.61$\pm$0.17 & 8.84$\pm$0.08 & 3870$\pm$683 & 3836$\pm$787 \\
56 & 2019/2/9 & 17:29:38 & 10.32$\pm$0.28 & 9.91$\pm$0.14 & 5563$\pm$1628 & 5601$\pm$1560 \\
57 & 2019/2/9 & 18:17:00 & 10.39$\pm$0.25 & 9.82$\pm$0.12 & 4646$\pm$1396 & 4647$\pm$1169 \\
58 & 2019/4/10 & 19:53:21 & 9.45$\pm$0.26 & 8.55$\pm$0.1 & 3496$\pm$886 & 3506$\pm$868 \\
59 & 2019/6/8 & 19:14:59 & 10.08$\pm$0.37 & 8.64$\pm$0.08 & 2503$\pm$608 & 2499$\pm$522 \\
60 & 2019/6/8 & 19:26:58 & 9.24$\pm$0.16 & 8.04$\pm$0.05 & 2864$\pm$302 & 2864$\pm$298 \\
61 & 2019/6/28 & 1:56:48 & 8.88$\pm$0.06 & 7.59$\pm$0.02 & 2717$\pm$98 & 2709$\pm$215 \\
62 & 2019/6/28 & 2:18:23 & 10.12$\pm$0.18 & 9.29$\pm$0.08 & 3688$\pm$647 & 3678$\pm$715 \\
63 & 2019/7/6 & 19:12:55 & 10.06$\pm$0.21 & 9.08$\pm$0.08 & 3301$\pm$588 & 3307$\pm$644 \\
64 & 2019/7/7 & 18:32:56 & 10.94$\pm$0.34 & 9.63$\pm$0.09 & 2686$\pm$654 & 2678$\pm$590 \\
65 & 2019/7/7 & 18:40:21 & 6.65$\pm$0.06 & 5.49$\pm$0.03 & 2934$\pm$122 & 2922$\pm$205 \\
66 & 2019/7/7 & 18:48:48 & 11.94$\pm$0.54 & 9.86$\pm$0.1 & 1876$\pm$519 & 1873$\pm$405 \\
67 & 2019/7/8 & 19:11:44 & 9.77$\pm$0.2 & 8.19$\pm$0.08 & 2333$\pm$257 & 2325$\pm$266 \\
68 & 2019/7/26 & 0:18:28 & 10.75$\pm$0.3 & 9.65$\pm$0.13 & 3047$\pm$799 & 3036$\pm$790 \\
69 & 2019/7/26 & 0:41:35 & 9.64$\pm$0.14 & 8.21$\pm$0.03 & 2517$\pm$193 & 2510$\pm$239 \\
70 & 2019/7/27 & 1:13:12 & 10.68$\pm$0.32 & 9.46$\pm$0.08 & 2830$\pm$680 & 2817$\pm$646 \\
71 & 2019/7/27 & 1:17:50 & 8.95$\pm$0.09 & 8.02$\pm$0.03 & 3420$\pm$240 & 3404$\pm$368 \\
72 & 2019/7/27 & 2:12:25 & 9.67$\pm$0.16 & 8.67$\pm$0.05 & 3256$\pm$399 & 3273$\pm$441 \\
73 & 2019/7/27 & 2:37:23 & 10.16$\pm$0.2 & 9.48$\pm$0.06 & 4183$\pm$932 & 4186$\pm$953 \\
74 & 2019/7/27 & 2:59:56 & 9.48$\pm$0.16 & 8.25$\pm$0.06 & 2813$\pm$295 & 2896$\pm$697 \\
75 & 2019/7/27 & 3:01:26 & 8.9$\pm$0.12 & 7.47$\pm$0.04 & 2517$\pm$170 & 2503$\pm$197 \\
76 & 2019/7/28 & 1:33:40 & 10.08$\pm$0.14 & 8.93$\pm$0.06 & 2952$\pm$290 & 2941$\pm$386 \\
77 & 2019/7/28 & 1:59:21 & 10.8$\pm$0.27 & 9.62$\pm$0.1 & 2899$\pm$598 & 2879$\pm$611 \\
78 & 2019/7/28 & 2:00:54 & 11.37$\pm$0.32 & 10.51$\pm$0.14 & 3603$\pm$1198 & 3592$\pm$1243 \\
79 & 2019/7/28 & 2:24:26 & 11.04$\pm$0.26 & 9.93$\pm$0.09 & 3027$\pm$626 & 3004$\pm$633 \\
80 & 2019/8/6 & 18:19:16 & 10.37$\pm$0.27 & 8.71$\pm$0.12 & 2245$\pm$340 & 2237$\pm$317 \\
81 & 2019/8/6 & 18:56:37 & 8.43$\pm$0.08 & 7.38$\pm$0.04 & 3148$\pm$189 & 3151$\pm$305 \\
82 & 2019/8/6 & 18:59:16 & 9.34$\pm$0.09 & 9$\pm$0.07 & 6104$\pm$1025 & 6149$\pm$1410 \\
83 & 2019/8/26 & 2:50:56 & 10.65$\pm$0.24 & 9.11$\pm$0.05 & 2379$\pm$309 & 2933$\pm$600 \\
84 & 2019/8/28 & 3:03:31 & 10.98$\pm$0.26 & 9.82$\pm$0.14 & 2934$\pm$638 & 2933$\pm$600 \\
85 & 2019/9/5 & 18:11:59 & 10.13$\pm$0.19 & 9.52$\pm$0.1 & 4466$\pm$1090 & 4443$\pm$1207 \\
86 & 2019/9/5 & 18:51:33 & 10.25$\pm$0.25 & 9.53$\pm$0.14 & 4038$\pm$1204 & 4016$\pm$1214 \\
87 & 2019/9/23 & 3:36:22 & 10.2$\pm$0.28 & 9.4$\pm$0.1 & 3777$\pm$1119 & 3777$\pm$1042 \\
88 & 2019/9/25 & 3:40:11 & 8.43$\pm$0.05 & 7.42$\pm$0.02 & 3234$\pm$120 & 3212$\pm$207 \\
89 & 2019/10/22 & 4:11:36 & 10$\pm$0.17 & 9.28$\pm$0.1 & 4038$\pm$802 & 4036$\pm$810 \\
90 & 2019/10/24 & 2:30:16 & 9.49$\pm$0.12 & 7.88$\pm$0.03 & 2299$\pm$138 & 2288$\pm$160 \\
91 & 2019/11/2 & 17:19:20 & 10.22$\pm$0.27 & 9.26$\pm$0.1 & 3348$\pm$842 & 3345$\pm$770 \\
92 & 2019/11/3 & 17:49:38 & 9.61$\pm$0.21 & 8.77$\pm$0.1 & 3659$\pm$782 & 3641$\pm$770 \\
93 & 2019/12/1 & 16:14:30 & 9.17$\pm$0.13 & 8.46$\pm$0.1 & 4073$\pm$652 & 4080$\pm$647 \\
\bottomrule
\end{tabular}
\end{table*}

\clearpage %
\begin{table}[htbp]
\centering
\begin{tabular}{@{}lcccccc@{}}
\toprule
Flash(ID) & Date(UT) & Time(UT) & R $\pm\ \sigma_R$(mag) & I $\pm\ \sigma_I$(mag) & T $\pm\ \sigma_T$(K) & T $\pm\ \sigma_T$(K) \citep{liakos2020neliota,liakos2024neliota} \\
\midrule
94  & 2019/12/1  & 16:23:14 & 8.42$\pm$0.06  & 5.57$\pm$0.04  & 1442$\pm$31   & 1438$\pm$53 \\
95  & 2019/12/1  & 16:30:43 & 10.75$\pm$0.24 & 9.12$\pm$0.13  & 2277$\pm$319  & 2277$\pm$353 \\
96  & 2019/12/1  & 17:14:41 & 11.16$\pm$0.36 & 9.35$\pm$0.13  & 2098$\pm$403  & 2093$\pm$363 \\
97  & 2019/12/20 & 4:34:17  & 9.5$\pm$0.19   & 8.51$\pm$0.06  & 3278$\pm$496  & 3254$\pm$477 \\
98  & 2020/1/30  & 17:18:09 & 11.03$\pm$0.31 & 9.51$\pm$0.11  & 2403$\pm$453  & 2393$\pm$427 \\
99  & 2020/1/30  & 17:35:39 & 10.62$\pm$0.19 & 9.75$\pm$0.11  & 3576$\pm$684  & 3558$\pm$727 \\
100 & 2020/3/1   & 16:54:24 & 8.32$\pm$0.06  & 7.15$\pm$0.02  & 2916$\pm$113  & 2919$\pm$161 \\
101 & 2020/3/1   & 17:10:06 & 9.92$\pm$0.25  & 9.42$\pm$0.12  & 5004$\pm$1477 & 4991$\pm$2002 \\
102 & 2020/3/27  & 17:40:25 & 10.01$\pm$0.12 & 8.7$\pm$0.04   & 2686$\pm$195  & 2677$\pm$263 \\
103 & 2020/3/29  & 18:14:11 & 10.83$\pm$0.2  & 9.72$\pm$0.08  & 3027$\pm$457  & 3030$\pm$507 \\
104 & 2020/3/29  & 19:16:47 & 10.18$\pm$0.19 & 9.26$\pm$0.06  & 3445$\pm$560  & 3430$\pm$605 \\
105 & 2020/4/28  & 19:19:55 & 8.99$\pm$0.07  & 8.13$\pm$0.02  & 3603$\pm$204  & 3587$\pm$323 \\
106 & 2020/6/25  & 18:28:18 & 7.92$\pm$0.07  & 6.66$\pm$0.03  & 2764$\pm$123  & 2763$\pm$192 \\
107 & 2020/6/26  & 19:52:32 & 9.73$\pm$0.07  & 8.22$\pm$0.02  & 2415$\pm$89   & 2408$\pm$160 \\
108 & 2020/7/26  & 19:08:21 & 10.2$\pm$0.09  & 9.13$\pm$0.1   & 3107$\pm$282  & 2788$\pm$365 \\
109 & 2020/7/26  & 19:10:25 & 9.15$\pm$0.04  & 7.82$\pm$0.02  & 2657$\pm$66   & 2649$\pm$171 \\
110 & 2020/8/13  & 0:57:11  & 10.16$\pm$0.24 & 9.03$\pm$0.07  & 2989$\pm$532  & 2991$\pm$540 \\
111 & 2020/8/14  & 0:54:21  & 9.29$\pm$0.12  & 8.53$\pm$0.05  & 3903$\pm$449  & 3910$\pm$623 \\
112 & 2020/8/14  & 1:15:36  & 9.95$\pm$0.18  & 9.45$\pm$0.1   & 5004$\pm$1242 & 4961$\pm$1518 \\
113 & 2020/12/9  & 3:09:58  & 9.83$\pm$0.17  & 9.32$\pm$0.09  & 4949$\pm$1159 & 4967$\pm$1227 \\
114 & 2021/3/17  & 17:46:53 & 9.48$\pm$0.1   & 7.97$\pm$0.03  & 2415$\pm$129  & 2410$\pm$168 \\
115 & 2021/3/17  & 18:07:19 & 10.47$\pm$0.21 & 9.21$\pm$0.06  & 2764$\pm$376  & 2754$\pm$391 \\
116 & 2021/4/18  & 19:13:59 & 9.43$\pm$0.13  & 7.89$\pm$0.04  & 2379$\pm$164  & 2370$\pm$184 \\
117 & 2021/4/18  & 20:19:24 & 9.74$\pm$0.19  & 8.64$\pm$0.08  & 3047$\pm$437  & 3038$\pm$460 \\
118 & 2021/5/15  & 18:38:41 & 10.12$\pm$0.2  & 9.38$\pm$0.08  & 3969$\pm$865  & 3971$\pm$926 \\
119 & 2021/5/18  & 20:08:22 & 9.9$\pm$0.19   & 8.66$\pm$0.06  & 2797$\pm$348  & 2788$\pm$365 \\
120 & 2021/6/15  & 19:06:13 & 9.96$\pm$0.13  & 9.19$\pm$0.06  & 3870$\pm$492  & 3863$\pm$596 \\
121 & 2021/6/15  & 19:23:19 & 10.86$\pm$0.28 & 10.19$\pm$0.16 & 4221$\pm$1391 & 4457$\pm$1492 \\
122 & 2021/6/15  & 19:38:52 & 8.15$\pm$0.05  & 6.73$\pm$0.02  & 2530$\pm$72   & 2520$\pm$158 \\
123 & 2021/7/6   & 2:11:11  & 10.82$\pm$0.08 & 10.33$\pm$0.15 & 5060$\pm$1082 & 5070$\pm$1586 \\
124 & 2021/7/15  & 19:22:29 & 10.2$\pm$0.26  & 8.76$\pm$0.07  & 2503$\pm$388  & 2497$\pm$385 \\
125 & 2021/7/16  & 18:49:32 & 9.96$\pm$0.33  & 9.15$\pm$0.11  & 3747$\pm$1263 & 3746$\pm$1274 \\
126 & 2021/8/2   & 1:26:06  & 10.41$\pm$0.28 & 9.85$\pm$0.13  & 4694$\pm$1488 & 4646$\pm$1967 \\
127 & 2021/8/2   & 1:34:28  & 8.75$\pm$0.16  & 8.28$\pm$0.08  & 5176$\pm$1169 & 5171$\pm$1433 \\
128 & 2021/8/2   & 2:44:36  & 10.12$\pm$0.25 & 9.72$\pm$0.15  & 5634$\pm$1595 & 5634$\pm$2105 \\
129 & 2021/8/2   & 2:51:03  & 8.85$\pm$0.15  & 7.74$\pm$0.04  & 3027$\pm$313  & 3022$\pm$349 \\
130 & 2021/10/3  & 3:14:17  & 10.17$\pm$0.17 & 9.37$\pm$0.07  & 3777$\pm$627  & 3791$\pm$726 \\
131 & 2021/10/11 & 16:57:00 & 8.66$\pm$0.08  & 7.73$\pm$0.03  & 3420$\pm$215  & 3404$\pm$385 \\
132 & 2021/10/12 & 16:31:16 & 9.33$\pm$0.21  & 8.25$\pm$0.08  & 3086$\pm$502  & 3066$\pm$516 \\
133 & 2021/10/12 & 17:42:18 & 9.75$\pm$0.17  & 8.75$\pm$0.07  & 3256$\pm$445  & 3248$\pm$528 \\
134 & 2021/12/8  & 16:15:11 & 8.58$\pm$0.09  & 7.59$\pm$0.04  & 3278$\pm$227  & 3261$\pm$322 \\
135 & 2021/12/8  & 16:34:22 & 10.17$\pm$0.3  & 8.2$\pm$0.05   & 1960$\pm$260  & 1953$\pm$252 \\
136 & 2022/4/5   & 17:30:56 & 8.87$\pm$0.15  & 7.53$\pm$0.05  & 2642$\pm$240  & 2635$\pm$265 \\
137 & 2022/4/5   & 17:54:38 & 9.31$\pm$0.13  & 7.95$\pm$0.05  & 2613$\pm$204  & 2600$\pm$232 \\
138 & 2022/6/3   & 18:21:31 & 7.96$\pm$0.09  & 6.64$\pm$0.02  & 2671$\pm$139  & 2666$\pm$192 \\
139 & 2022/6/4   & 18:20:51 & 9.4$\pm$0.27   & 8.32$\pm$0.08  & 3086$\pm$671  & 3081$\pm$624 \\
\bottomrule
\end{tabular}
\end{table}

\clearpage %
\begin{table}[htbp]
\centering
\begin{tabular}{@{}lcccccc@{}}
\toprule
Flash(ID) & Date(UT) & Time(UT) & R $\pm\ \sigma_R$(mag) & I $\pm\ \sigma_I$(mag) & T $\pm\ \sigma_T$(K) & T $\pm\ \sigma_T$(K) \citep{liakos2020neliota,liakos2024neliota} \\
\midrule
140 & 2022/6/4   & 18:22:47 & 10.15$\pm$0.38 & 9.19$\pm$0.14  & 3347$\pm$1229 & 3326$\pm$1077 \\
141 & 2022/6/4   & 19:44:17 & 10.06$\pm$0.29 & 9.45$\pm$0.14  & 4466$\pm$1473 & 4477$\pm$1909 \\
142 & 2022/6/23  & 1:46:08  & 8.73$\pm$0.12  & 7.22$\pm$0.03  & 2415$\pm$154  & 2409$\pm$207 \\
143 & 2022/7/22  & 2:13:27  & 9.97$\pm$0.23  & 8.73$\pm$0.06  & 2797$\pm$427  & 2788$\pm$425 \\
144 & 2022/7/22  & 2:48:11  & 10.27$\pm$0.35 & 9.29$\pm$0.13  & 3301$\pm$1114 & 3297$\pm$993 \\
145 & 2022/7/22  & 2:49:51  & 9.36$\pm$0.28  & 7.51$\pm$0.09  & 2062$\pm$281  & 2056$\pm$257 \\
146 & 2022/8/1   & 18:27:23 & 9.93$\pm$0.25  & 8.66$\pm$0.12  & 2748$\pm$498  & 2744$\pm$516 \\
147 & 2022/8/1   & 18:28:19 & 9.42$\pm$0.18  & 7.62$\pm$0.1   & 2107$\pm$197  & 2435$\pm$410 \\
148 & 2022/8/3   & 18:23:43 & 9.76$\pm$0.08  & 8.37$\pm$0.05  & 2571$\pm$132  & 2575$\pm$204 \\
149 & 2022/8/4   & 18:47:07 & 9.26$\pm$0.16  & 7.63$\pm$0.11  & 2277$\pm$217  & 2271$\pm$242 \\
150 & 2022/9/1   & 18:33:36 & 10.66$\pm$0.27 & 8.04$\pm$0.05  & 1549$\pm$141  & 1543$\pm$186 \\
151 & 2022/10/19 & 3:03:44  & 8.47$\pm$0.08  & 8.25$\pm$0.04  & 7372$\pm$1028 & 7426$\pm$1907 \\
152 & 2022/10/20 & 1:12:58  & 8.58$\pm$0.1   & 7.38$\pm$0.02  & 2864$\pm$178  & 2858$\pm$285 \\
153 & 2022/10/20 & 2:56:38  & 9.08$\pm$0.08  & 8.59$\pm$0.04  & 5060$\pm$539  & 5084$\pm$755 \\
154 & 2022/10/21 & 2:55:12  & 10.22$\pm$0.22 & 9.56$\pm$0.07  & 4260$\pm$1071 & 4254$\pm$1213 \\
155 & 2022/10/22 & 3:07:57  & 10.92$\pm$0.23 & 10.27$\pm$0.13 & 4299$\pm$1231 & 4285$\pm$1532 \\
156 & 2022/10/22 & 3:25:18  & 10.44$\pm$0.06 & 9.55$\pm$0.11  & 3522$\pm$343  & 3512$\pm$466 \\
157 & 2022/10/22 & 3:39:55  & 10.7$\pm$0.06  & 9.91$\pm$0.03  & 3807$\pm$210  & 3800$\pm$543 \\
158 & 2022/10/22 & 3:55:36  & 9.51$\pm$0.12  & 8.68$\pm$0.05  & 3688$\pm$396  & 3677$\pm$488 \\
159 & 2022/10/22 & 3:56:50  & 10.31$\pm$0.24 & 9.51$\pm$0.09  & 3777$\pm$954  & 3778$\pm$935 \\
160 & 2022/10/29 & 17:03:58 & 9.25$\pm$0.12  & 7.72$\pm$0.03  & 2391$\pm$150  & 2382$\pm$279 \\
161 & 2022/10/30 & 16:41:06 & 10.08$\pm$0.28 & 9.26$\pm$0.12  & 3717$\pm$1120 & 3711$\pm$1109 \\
162 & 2022/10/30 & 16:47:28 & 9.68$\pm$0.19  & 8.56$\pm$0.05  & 3008$\pm$402  & 3005$\pm$472 \\
163 & 2022/10/30 & 16:54:55 & 9.65$\pm$0.16  & 8.57$\pm$0.05  & 3086$\pm$356  & 3080$\pm$453 \\
164 & 2022/10/30 & 17:34:16 & 8.64$\pm$0.08  & 8.03$\pm$0.05  & 4466$\pm$427  & 4475$\pm$883 \\
165 & 2022/10/31 & 19:17:17 & 8.6$\pm$0.16   & 7.51$\pm$0.05  & 3067$\pm$349  & 3065$\pm$517 \\
166 & 2022/11/19 & 2:39:28  & 8.82$\pm$0.13  & 7.92$\pm$0.05  & 3496$\pm$382  & 3492$\pm$462 \\
167 & 2022/12/18 & 3:37:48  & 9.11$\pm$0.13  & 7.74$\pm$0.04  & 2599$\pm$197  & 2577$\pm$225 \\
168 & 2022/12/26 & 15:46:17 & 7.76$\pm$0.07  & 6.38$\pm$0.05  & 2585$\pm$121  & 2579$\pm$197 \\
169 & 2022/12/26 & 16:48:14 & 10.26$\pm$0.17 & 9.68$\pm$0.08  & 4600$\pm$1004 & 4580$\pm$1240 \\
170 & 2022/12/27 & 16:18:50 & 9.64$\pm$0.2   & 9.08$\pm$0.11  & 4694$\pm$1239 & 4669$\pm$1310 \\
171 & 2022/12/27 & 17:47:37 & 9.29$\pm$0.14  & 7.98$\pm$0.05  & 2686$\pm$232  & 2675$\pm$298 \\
172 & 2022/12/27 & 18:11:32 & 9.73$\pm$0.04  & 8.53$\pm$0.1   & 2864$\pm$188  & 2856$\pm$307 \\
173 & 2023/1/16  & 4:11:21  & 10.09$\pm$0.25 & 9.04$\pm$0.08  & 3148$\pm$641  & 3144$\pm$623 \\
174 & 2023/2/22  & 17:49:41 & 10.37$\pm$0.15 & 10.15$\pm$0.1  & 7372$\pm$1406 & 7336$\pm$2244 \\
175 & 2023/2/22  & 17:52:04 & 9.97$\pm$0.12  & 8.7$\pm$0.03   & 2748$\pm$200  & 2745$\pm$381 \\
176 & 2023/3/26  & 20:25:29 & 9.88$\pm$0.15  & 8.79$\pm$0.05  & 3067$\pm$328  & 3054$\pm$428 \\
177 & 2023/4/23  & 18:04:40 & 10.6$\pm$0.18  & 9.61$\pm$0.09  & 3278$\pm$503  & 3260$\pm$600 \\
178 & 2023/4/24  & 17:58:31 & 10.47$\pm$0.31 & 9.1$\pm$0.07   & 2599$\pm$518  & 2597$\pm$485 \\
179 & 2023/4/24  & 20:02:42 & 9.63$\pm$0.22  & 8.28$\pm$0.06  & 2627$\pm$353  & 2630$\pm$400 \\
180 & 2023/5/23  & 20:06:15 & 8.94$\pm$0.09  & 8.16$\pm$0.04  & 3839$\pm$320  & 3860$\pm$681 \\
181 & 2023/5/24  & 20:11:10 & 8.32$\pm$0.05  & 6.41$\pm$0.01  & 2010$\pm$43   & 2824$\pm$237 \\
182 & 2023/5/24  & 21:03:20 & 9.1$\pm$0.09   & 7.53$\pm$0.02  & 2344$\pm$106  & 2335$\pm$262 \\
183 & 2023/5/25  & 20:34:28 & 7.51$\pm$0.03  & 6.28$\pm$0.01  & 2813$\pm$52   & 2935$\pm$465 \\
184 & 2023/5/26  & 18:17:44 & 9.79$\pm$0.37  & 8.25$\pm$0.08  & 2379$\pm$538  & 2376$\pm$469 \\
185 & 2023/6/21  & 18:47:44 & 9.21$\pm$0.06  & 9$\pm$0.04     & 7505$\pm$914  & 7520$\pm$2682 \\
186 & 2023/6/22  & 19:58:09 & 11$\pm$0.29    & 10.24$\pm$0.13 & 3903$\pm$1247 & 3911$\pm$1643 \\
187 & 2023/6/22  & 20:03:20 & 10.18$\pm$0.14 & 8.53$\pm$0.03  & 2256$\pm$155  & 2372$\pm$375 \\
\bottomrule
\end{tabular}
\end{table}

\clearpage %
\bibliographystyle{jasr-model5-names}
\biboptions{authoryear}
\bibliography{jasr-template}

\end{document}